\documentclass[preprint2]{emulateapj}

\newcommand{\chandra}{\emph{Chandra}}
\newcommand{\fe}{Fe}
\newcommand{\ka}{K$\alpha$}
\newcommand{\kb}{K$\beta$}

\slugcomment{Accepted version 1.00, Apr 12th 2010}

\shorttitle{Fluorescence Fe lines with \chandra\ }
\shortauthors{Torrej\'on et al.}

\begin{document}


\title{A {\it Chandra} survey of fluorescence Fe lines in X-ray
  Binaries at high resolution}


\author{J.M. Torrej\'on\altaffilmark{1,2},
  N.S. Schulz\altaffilmark{2}, M.A. Nowak\altaffilmark{2} and
  T.R. Kallman\altaffilmark{3}}


\altaffiltext{1}{Instituto de F\'isica Aplicada a las Ciencias y las Tecnolog\'ias, Universidad de Alicante, E 03080 Alicante, Spain; jmt@ua.es}  
\altaffiltext{2}{MIT Kavli Institute for Astrophysics and Space
  Research, Cambridge MA 02139} 
\altaffiltext{3}{NASA Goddard Space Flight Center, Greenbelt, MD 20771}

\begin{abstract}

\fe\ K line fluorescence is commonly observed in the X-ray spectra of
many X-ray binaries and represents a fundamental tool to investigate
the material surrounding the X-ray source.  In this paper we present a
comprehensive survey of 41 X-ray binaries (10 HMXBs and 31 LMXBs) with \chandra\, with specific emphasis on the
Fe K region and the narrow \fe\ \ka\  line, at the highest resolution possible. We find that: {\it a}) The \fe\ \ka\ line is always centered at $\lambda=1.9387\pm 0.0016$ \AA, compatible with Fe \textsc{i} up to Fe \textsc{x}; we detect no shifts to higher ionization states nor any difference between HMXBs and LMXBs.  {\it b}) The line is very narrow, with $FWHM\leq 5$ m\AA, normally not resolved by \chandra\ which means that the reprocessing material is not rotating at high speeds.  {\it c}) \fe\ \ka\ fluorescence is present in all the HMXB in the survey. In
contrast, such emissions are astonishingly rare ($\sim 10$ \% ) among low mass X-ray
binaries (LMXB) where only a few out of a large number showed \fe\ K
fluorescence. However, the line and edge properties of these few are
very similar to their high mass cousins.  {\it d}) The lack of Fe line emission is always accompanied by the lack of any detectable K edge.  {\it e}) We obtain the empirical curve
of growth of the equivalent width of the \fe\ \ka\ line versus the density column of the reprocessing material, i.e. $EW_{\rm K\alpha}$ vs $N_{\rm H}$, and show that it is
consistent with a reprocessing region spherically distributed around
the compact object.  {\it f}) We show that
fluorescence in X-ray binaries follows the X-ray Baldwin effect as
previously only found in the X-ray spectra of active galactic nuclei. We interpret this finding as evidence of decreasing neutral Fe abundance with increasing X-ray illumination and use it to explain some spectral states of Cyg X-1 and as a possible cause of the lack of narrow Fe line emission in LMXBs.  {\it g}) Finally, we study 
anomalous morphologies such as Compton shoulders and asymmetric line
profiles associated with the line fluorescence. Specifically, we
present the first evidence of a Compton shoulder in the HMXB
X1908+075.  Also the \fe\ \ka\ lines of 4U1700$-$37 and LMC X-4
present asymmetric wings suggesting the presence of highly structured
stellar winds in these systems.

\end{abstract}


\section{Introduction}

\fe\ K fluorescence lines constitute a fundamental tool to probe the
physical characteristics of the material in the close vicinity of
X-ray sources \citep{george91}.  In the spectra of accreting X-ray
Binaries (XRBs) these lines are very prominent due to of their
intrinsic X-ray brightness and ubiquitous stellar material.  High mass X-ray binaries (HMXB) are
composed of a compact object, either a neutron star (NS) or a stellar
size black hole (BH), accreting from the powerful wind of a massive OB
type star. The compact object in these cases is deeply embedded into
the stellar wind of the donor providing an excellent source of
illumination. In contrast, in low mass (LMXBs) and intermediate mass
X-ray binaries (IMXBs), the donor stars are not significant sources of stellar winds. However they usually feature strong and matter rich
accretion disks around the orbiting compact object and also have
associated outflow processes which, in turn, tend to be not so
important in HMXB.

\begin{figure}
\includegraphics[angle=-90,width=\columnwidth]{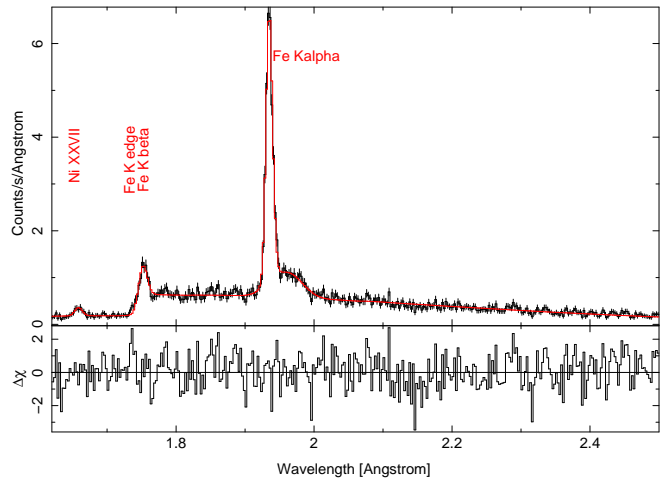}
\caption{{\it Chandra HETG} spectra of the HMXB GX301$-$2, included in this survey, ObsID 2733, showing all the relevant features discussed in the present work: \fe\ \ka\ and  \fe\ \kb\ fluorescence lines, Fe K edge, Compton shoulder and a hot line.}
\label{fig:fecomplex}
\end{figure}

Fluorescence excitation occurs whenever there is a low ionization gas illuminated by X-rays. In XRBs, the strong point like source of X-rays, allows to observe strong fluorescence emission.  The X-ray source, powered via accretion,
irradiates the 
circumstellar material, either the wind or the accretion disk.  Whenever this material is more neutral than Li-like, the Fe atoms present
in the stellar wind absorb a significant fraction of continuum photons
blueward of the K edge (at $\sim 1.74$ \AA) thereby removing K shell
electrons.  The vacancy thus produced will be occupied by electrons
from the upper levels producing K$\alpha$ (L $\rightarrow$ K) and
K$\beta$ (M $\rightarrow$ K) fluorescence emission lines at $\sim$
1.94 \AA\ and 1.75 \AA\ respectively (Fig. \ref{fig:fecomplex}). K-shell fluorescence
emission is highly inefficient for electron numbers $Z\leq 16$.  The \emph{Auger} effect dominates at lower $Z$
although K shell emission can be observed under very specific
circumstances~\citep{schulz02}. However, the fluorescence yield increases monotonically with $Z$. In the case of \fe,
fluorescence yields are already quite competitive \citep[0.37, ][]{palmeri03}. Furthermore, the Fe is abundant and appears in an unconfused part of the spectrum. Therefore \fe\ K fluorescence is observed in a wide range of objects. 

The \fe\ \ka\ line can present a composite structure. A broad line
component, with {\it FWHM} of the order of keV and a narrow line
component with {\it FWHM} much lower, of the order of some eV
\citep{miller02, hanke09}.  However, as has been shown in \cite{hanke09}, the detection of the broad component by \chandra\ is very difficult and requires simultaneous {\it RXTE} coverage. Furthermore, Nowak et al. (in prep.), observing Cyg X-1 simultaneously with several X-ray telescopes, have shown that while {\it Suzaku} and {\it RXTE} show clearly the broad component, \chandra\ does not. On the other hand, {\it Suzaku} agrees on the narrow component detected by \chandra. In the present survey, we will focus specifically on the narrow component, which is best studied at high resolution.  \chandra\ high energy
transmission gratings (HETG, \cite{canizares05}) are very well suited for this
purpose. While {\it RGS} instrument on board {\it XMM-Newton} has the
required spectral resolution, it lacks of effective area shortward of 6 \AA.


The photons emitted during fluorescence must further travel through
the stellar wind to reach the interstellar medium. In some cases,
these photons can be Compton downscattered to lower energies and a
'red shoulder' can be resolved in the Fe line (\cite{watanabe03}).  In
such a case, the Compton shoulder can be used as a further probe of
the wind material.

\cite{gott95} established a comprehensive catalog of Fe line sources
using \emph{EXOSAT} GSPC.  These authors were able to detect iron line
emission in 51 sources out of which 32 were identified as X-ray
binaries (XRB). From these, 20 ($\sim 63\%$) were LMXB and 12 ($\sim
37\%$) HMXB. On average, the former showed a broad ($\sim 1$ keV) line
centered at $6.628\pm 0.012$ keV, while the latter tended to show
narrower ($\sim 0.5$ keV) lines centered at $6.533\pm 0.003$ keV. \emph{EXOSAT} GSPC had a spectral resolution of about this amount and, therefore,  a width of $\sim 0.5$ keV represents an upper limit.  More
recently, \cite{asai00} have performed a study of the Fe K line in a
sample of 20 LMXB using ASCA GIS and SIS data. These authors were able
to detect significant Fe line emission in roughly half of the
sources. This line tended to be centered around 6.6 keV but showed
large scatter with extreme values going from $\sim 6.55$ to 6.7 keV. In
general, the FWHM is not resolved but for those sources where the
width could be measured was $\sim 0.5$ keV.

In this paper we study in a homogeneous way and at the highest
spectral resolution, the narrow component of the Fe line for the whole
sample of HMXB and LMXB currently public within the \chandra\ archive.
Specific studies of some individual sources within our sample have
been published elsewhere (e.g. \cite{watanabe06} for Vela X-1). However,
we have reprocessed the entire sample to guarantee its homogeneity and
have reanalyzed it, focused specifically on the narrow component.

\section{Observations}

\begin{figure*}
\includegraphics[angle=-90,width=\columnwidth]{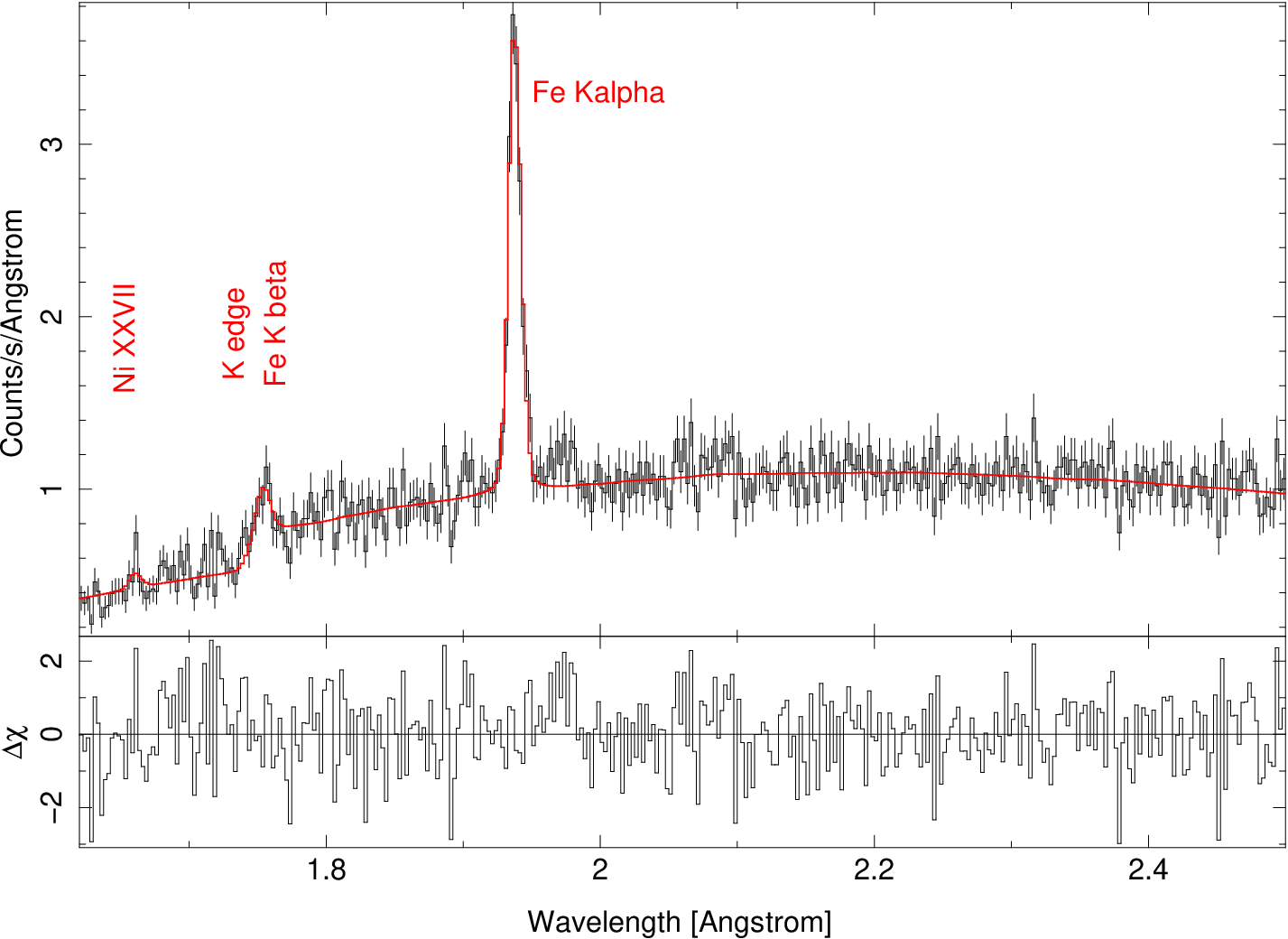}
\includegraphics[angle=-90,width=\columnwidth]{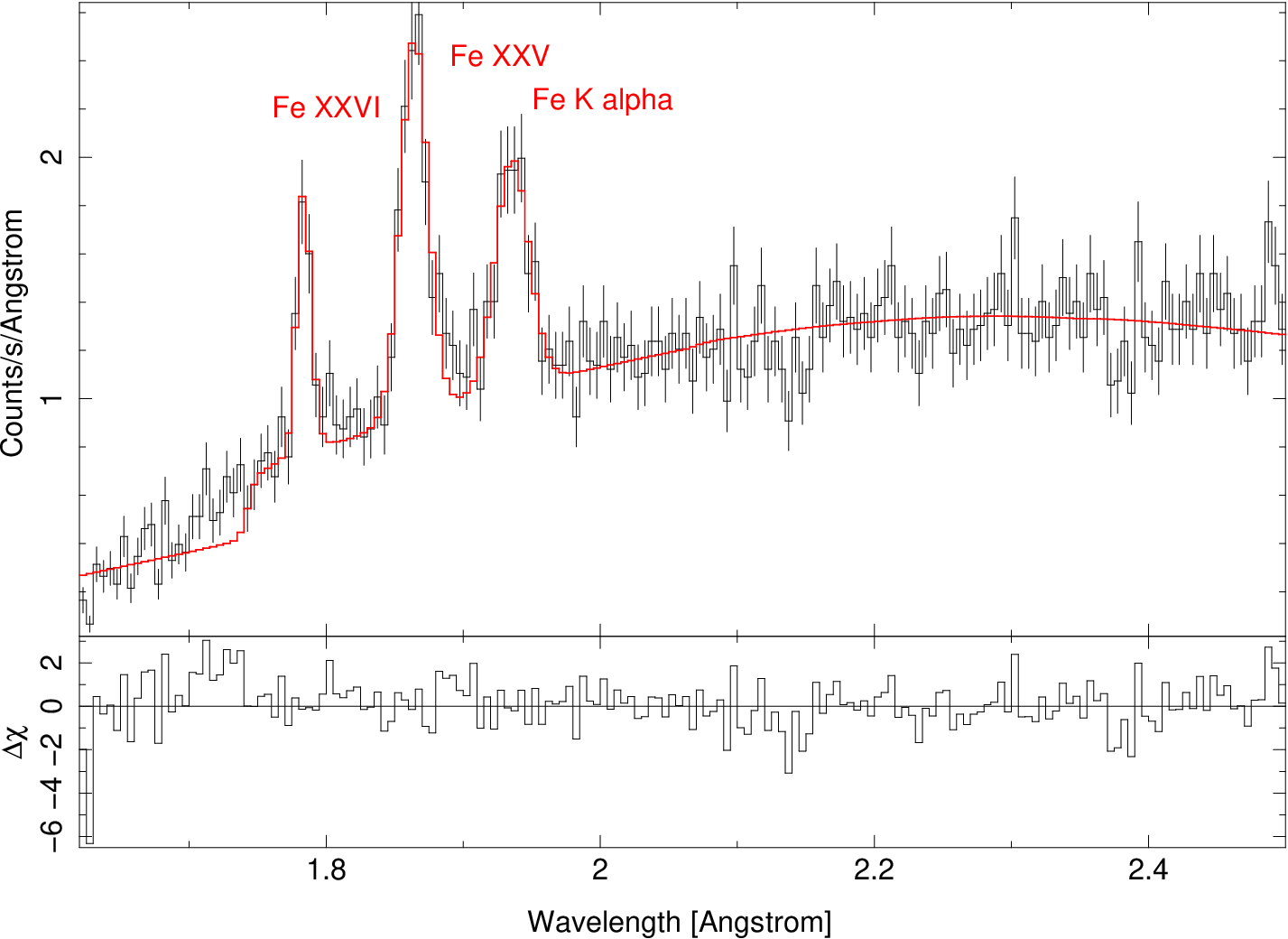}

\includegraphics[angle=-90,width=\columnwidth]{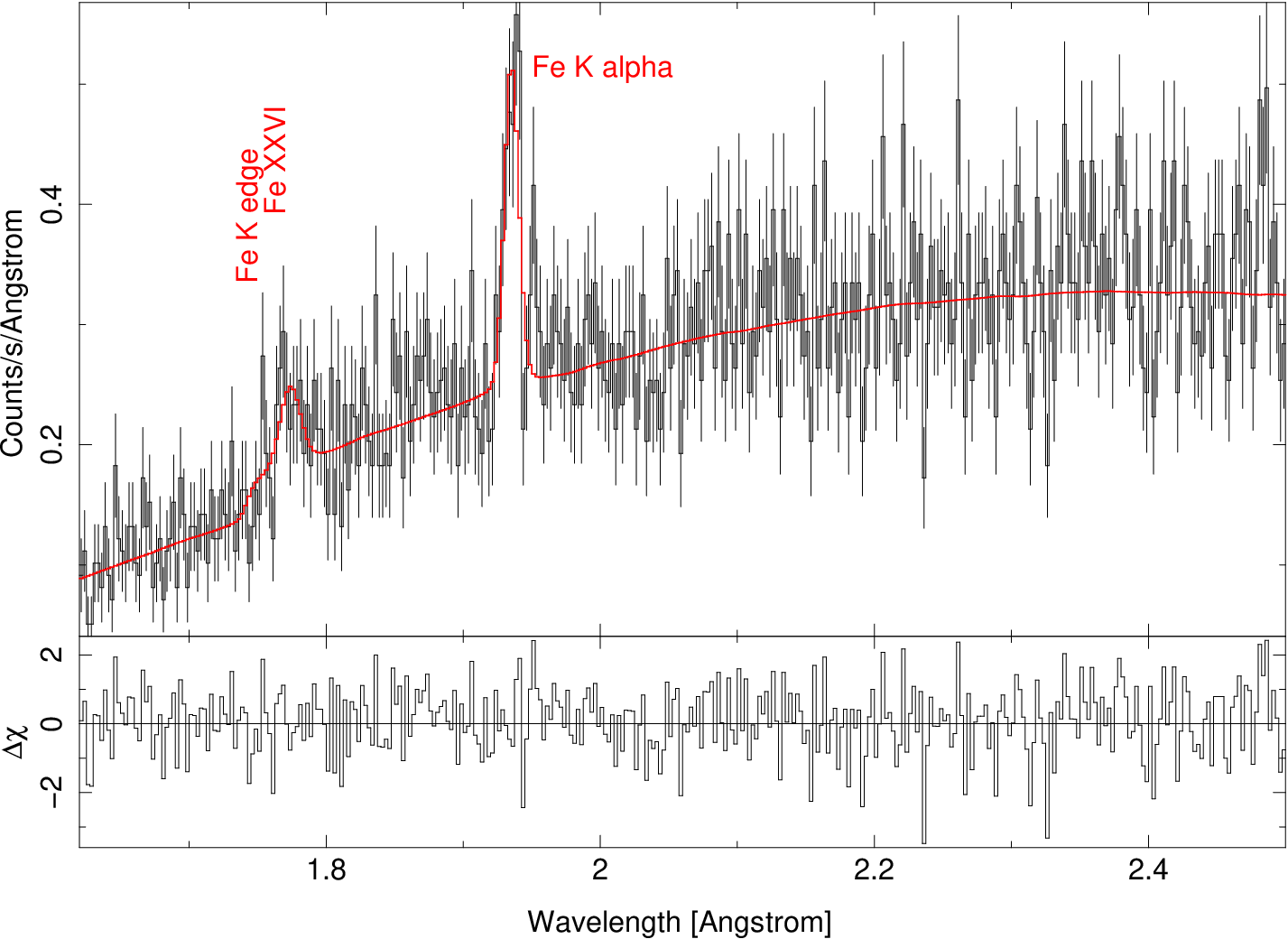}
\includegraphics[angle=-90,width=\columnwidth]{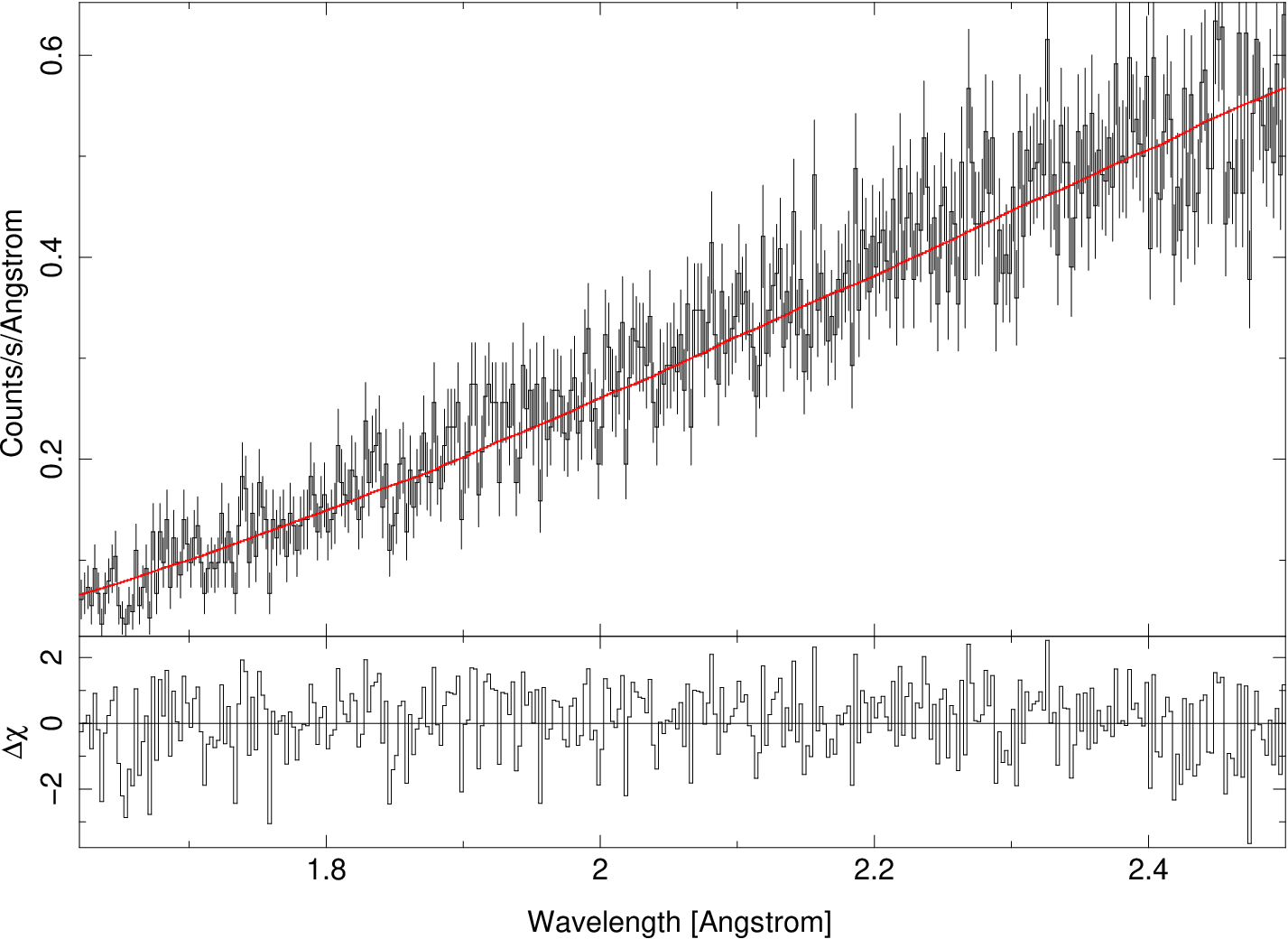}
\caption{Four representative examples of the fitting process. Upper row: HMXBs. Left: Vela X-1 shows Fe\ \ka\ and Fe\ \kb\ but not hot photoionization lines (except Ni \textsc{xxvii}). Right: Cyg X-3, in turn, shows hot photoionization lines of Fe \textsc{xxv} and Fe \textsc{xxvi} which, in low resolution and broadband spectra, are confused with the true cold Fe\ \ka\ fluorescence line. Lower row: LMXBs. Left: 4U1822$-37$ shows Fe\ \ka\ and a hot line. Right: The spectra of 4U1957$+$11, as the vast majority of LMXBs, shows no lines at all, in this case superimposed on a pure powerlaw continuum.}
\label{fig:fitting}
\end{figure*}

We have reprocessed all the available HETGS data for OB stellar
systems which in the end involved 10 sources. We also searched for
\fe\ fluorescence emission from 31 LMXB and found 4 cases of late type
companions with positive detections. The sample is presented in Table
\ref{tab:sample}. In the case of HMXBs we included observational
information of all 10 sources available since we detected line
fluorescence in all candidates. In the case of LMXBs we include such
information only for those with clear detections, but list all the
other targets in the footnote for reference purposes.

\begin{deluxetable*}{llccccl}
\tabletypesize{\scriptsize}
\tablecaption{The complete sample of X-ray binaries analyzed in this work.}
\tablewidth{0pt}
\tablehead{
\colhead{Source} & \colhead{Alternative name} & \colhead{$\alpha$} & \colhead{$\delta$} & \colhead{MK donor} & \colhead{$d$(kpc)}   & \colhead{ObsID}
}

\startdata
\cutinhead{HMXB}
Cen X-3       & 4U 1119-603  & 11 21 15.78 & -60 37 22.7 & O6.5II-III & 8  & 705, 1943\\
OAO1657$-$415 & EXO 1657-419 & 17 00 47.90  & -41 40 23   & Ofpe  &  7.1$\pm$1.3 & 1947 \\
Cyg X-1       & 4U 1956+35   & 19 58 21.67 & +35 12 05.77 & O9.7Iab        & 2.15$\pm$ 0.07  & 2415, 3815 \\
Cyg X-3       & 4U 2030+40 & 20 32 25.78 & +40 57 27.9 & WR?      & 9  & 425, 426, 1456\\
X1908$+$075   & 4U 1909+07 & 19 10 48 & +07 35.9 & O7.5-9.5If & 7 & 5476, 5477, 6336\\
Vela X-1      & 4U 0900-40 & 09 02 06.86 & -40 33 16.90 & B0Iab       & 1.9$\pm$0.1 & 102, 1926, 1927, 1928 \\
4U1700$-$37   &            & 17 03 56.77 & -37 50 38.91 & O6.5Iaf     & 1.7 & 657\\
GX301$-$2     & 4U 1223-62 & 12 26 37.60  & -62 46 14 & B1.5Ia+        & 3$^{+1}_{-0.5}$ & 103, 2733, 3433\\
LMC X-4       & 4U 0532-66 & 05 32 49.79 & -66 22 13.8 &  O8III       & 50  & 9571, 9573, 9574\\
$\gamma$ Cas$^{a}$  & 4U 0054+60 & 00 56 42.53 & +60 43 00.26 & B0.5IIIe   & 0.19 & 1985 \\
\cutinhead{LMXB}
4U1822-371 &           &  18 25 46.8 & -37 06 19 & M0V &    & 671 \\
GX1+4 & 4U 1728-24 & 17 32 02.16 & -24 44 44.02 & M5III &    &  2710 \\
Her X-1 & 4U 1656+35 & 16 57 49.83 & +35 20 32.6 & A5V & 4.5 & 2749, 3821, 3822, 4585 \\
& & & & & & 6149, 3821, 6150\\
Cir X-1 & 4U 1516-56   & 15 20 40.874 & -57 10 00.26 & B5-A0  &   & 706, 1700\\
\enddata
\tablecomments{Only sources with positive detections have been quoted above. E: Eclipsing binary, P: X-ray pulsar. Sources analyzed with negative detections: LMXB: 2S0918-549, 2S0921-63, 4U1254-690, 4U1543-62, 4U1624-49, 4U1626-67, 4U1636-53, 4U1705-44, 4U1728-16, 4U1728-34, 4U1735-44, 4U1820-30, 4U1822-00, 4U1916-053, 4U1957+11, 4U2127+119, Cyg X-2, EXO0748-676, GRO J1655-40, GRS 1747-312, GRS 1758-258, GX13+1. GX17+2, GX3+1, GX339-4, GX340+0, GX349+2, GX5-1, GX9+1, Ginga1826-238, SAX J1747.0-2853, SAX J1808.4-3658, ScoX-1, Ser X-1, XTE J1118+480, XTE J550-564, XTE J1650-500, XTE J 1746, XTE J1814-338. HMXB: Cyg X-1 ObsIDs 107, 1511, 2741, 2742, 2743, 3407, 3724, Cir X-1 ObsIDs 1905, 1906, 1907, 5478, 6148. }
\tablenotetext{a}{Status as a HMXB still unclear. Not included in the correlations.}
\label{tab:sample}
\end{deluxetable*}

There were a few peculiar sources, such as SS 433 or $\gamma$Cas, in
which the \fe\ K region is dynamically either too complex or where the
nature of the source as an accreting source is fundamentally in
question. Some sources, like LS5039, LMC X-1 and LMC X-3, were
excluded because there was no significant X-ray emission in the
K$\alpha$ region under study. There are many observations of Cyg X-1
in the archive, but only two of these showed positive detections of
the \fe\ka\ line. In this case we list only the ObsIDs where we
detected line fluorescence. Likewise, sources presenting \emph{only} warm absorption lines, like 4U1624$-$49 have also been excluded.

For the present study we focus on the 1.6 - 2.5 \AA\ ($\approx 4.96$
keV - 7.74 keV) spectral region, which contains both Fe K$\alpha$, Fe
K$\beta$ emissions and the Fe K edge (Fig.\ref{fig:fecomplex}). Pile up has been found to be
negligible in this spectral region except for very few sources not
included in this survey. We treat the spectral continuum as local and
use a simple powerlaw modified either by an edge at 1.740 \AA\ or a
photoelectric absorption (\texttt{phabs, tbabs}).  For the latter case
we apply solar abundances from \citet{anders89} and cross sections
from \citet{balun92}.  Our focus on the local continuum is justified by the fact that the process of inner K shell line fluorescence of neutral
matter has a fundamentally different relation to its 
underlying continuum than in the photoionization of warm
plasmas. While the latter requires photoionization 
energies much lower than the K edge energy to remove 
electrons down to He-like ions
of Fe, it requires photon energies beyond the Fe K edge
of 7.1 keV to remove an electron from the K shell of neutral
Fe. In this respect, Fe K shell fluorescence is entirely 
independent of the nature of the X-ray continuum below the 
Fe K edge ($\lambda > 1.74$ \AA) which allows us to focus on local continua
only. Furthermore, the recovery of the continuum beyond the
K edge ($\lambda < 1.74$ \AA ) goes with the power of $\sim 3$ and is, thus, extremely steep,
requiring to account for the bulk of absorption of 
photons only very close to the edge itself. This provides us with both the optical depth of
the edge $\tau_{\rm edge}$ and the equivalent column density of the
reprocessing material, $N_{\rm H}$. Given the spectral range we have
focused on, this latter quantity is measured from the K edge via the
assumed abundance of Fe with respect to H.  Apart from Fe K$\alpha$
and K$\beta$ emissions, other emission lines from photoionised plasma
have been detected in many cases, most prominently at 1.85 \AA\ ($\approx 6.7$
keV, Fe \textsc{xxv}), 1.78 \AA\ ($\approx 6.96$
keV, Fe \textsc{xxvi} Ly $\alpha$), and 1.66 \AA\ ($\approx 7.46$
keV, Ni
\textsc{xxvii} Ly $\alpha$). Whenever present, these lines were fitted
with Gaussians, to get a good fit in the whole wavelength range, but we do not study them here because they are out of the scope of the present paper. The edge has been fixed at 1.740 \AA\ (=7.125 keV)
for all the sources\footnote{except for LMC X-4 ObsId 9574, where the
  analysis of highly ionized lines indicated that it could be
  blueshifted to 1.72 \AA (=7.208 keV) pointing to a slightly higher
  ionization degree (J. Lee, priv. comm.)}. 

In Fig. \ref{fig:fitting} we present four representative examples of the quality of the fits. The upper row shows two HMXBs of our sample, Vela X-1 (left) and Cygnus X-3 (right). As can be seen, the spectra of Vela X-1 is relatively clean, showing only Fe\ \ka, Fe\ \kb\ and the K edge (plus a small hot line beyond the edge). The spectra of Cygnus X-3, in turn, is more complex. It shows two hot lines, Fe \textsc{xxv} and Fe \textsc{xxvi} of photoionized iron. These latter lines, in particular Fe \textsc{xxv}, can not be resolved in low resolution and broad band spectra and are often confused with the true fluorescence Fe\ \ka\ of (near)neutral iron. Hence the paramount importance of using \chandra\ for the present survey. The lower row corresponds to two LMXBs within our sample. The case of 4U1822$-$37 (left) shows a morphology similar to that of the HMXBs. The high resolution now enables us to discern the hot line Fe \textsc{xxvi}, at 1.78 \AA, of photoionized iron not confusing it with Fe\ \kb\ at 1.75 \AA. Finally, we show the example of 4U1957$+$11 (right) which, as the vast majority of LMXBs, does not show any emission line or edge  at all. The reduced $\chi^{2}$ is $\approx 1$ in all cases.

The data were rebinned to
match the resolution of HEG (0.012 \AA\ {\it FWHM}) and MEG (0.023
\AA\ {\it FWHM}). Except in a few cases, the line width was not or barely
resolved. In those cases we fixed the line width to $\sigma=0.005$
\AA. The measured $EW$, though, were found to be independent of this
value.

In some sources, a Compton shoulder redward of \ka\ can be
discerned. In those cases the shoulder was modeled using Gaussian
functions. Shoulder properties are presented in Table \ref{tab:compt}.

Both HEG and MEG data were analyzed. However, only HEG data were used
to derive the line properties. First orders ($m=\pm 1$) were fitted
simultaneously and MEG data were checked in all cases for
consistency. In cases of low S/N the MEG data were also used to reduce
errors.  All analyzed sources are listed in Table
\ref{tab:sample}. The spectra and the response files (arf and rmf)
were extracted using standard CIAO software (v 4.4) and analyzed with
Interactive Spectral Interpretation System (ISIS) v 1.4.9-19 \citep{houck02}.

\begin{figure*}
\includegraphics[angle=-90,width=\columnwidth]{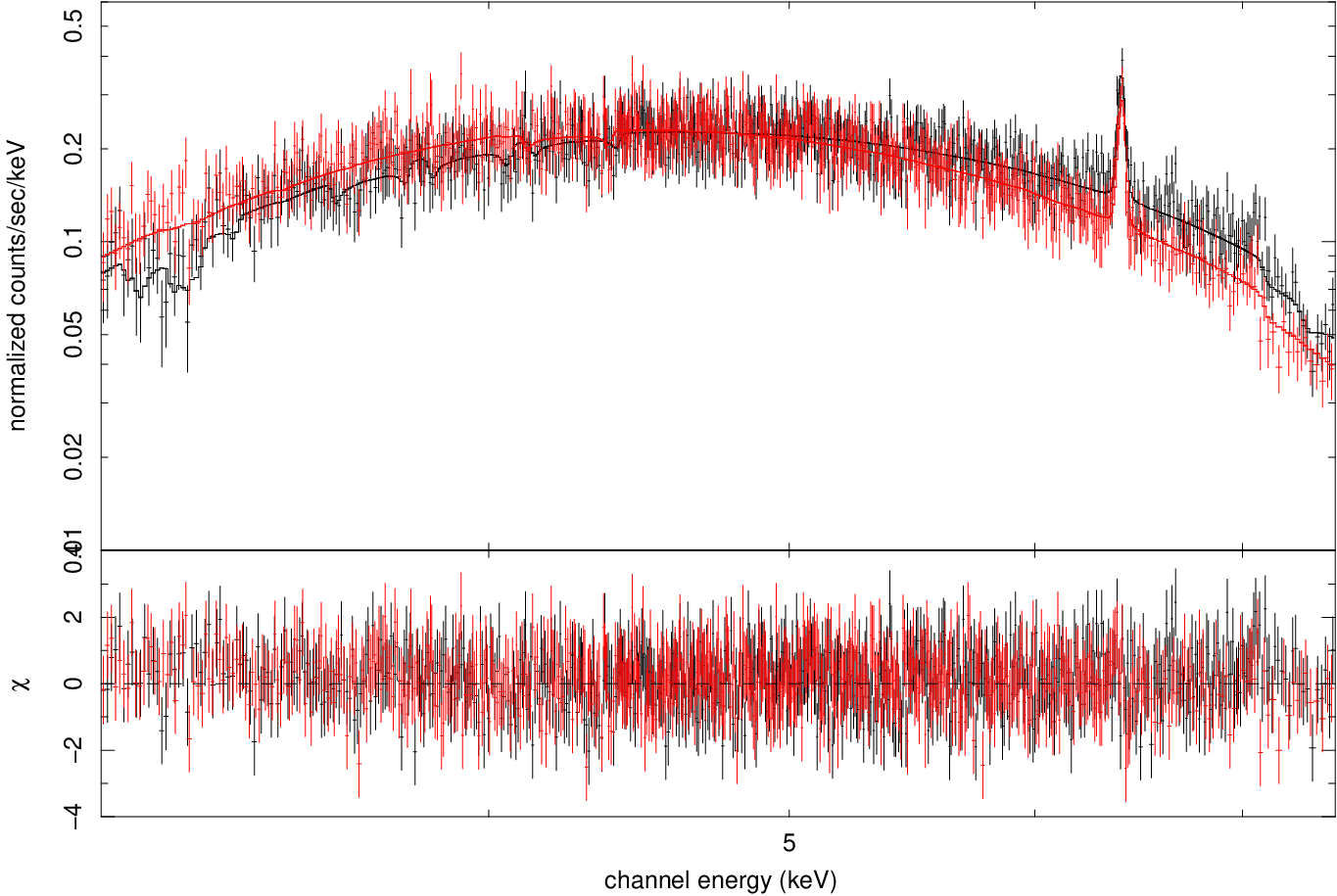}
\includegraphics[angle=-90,width=\columnwidth]{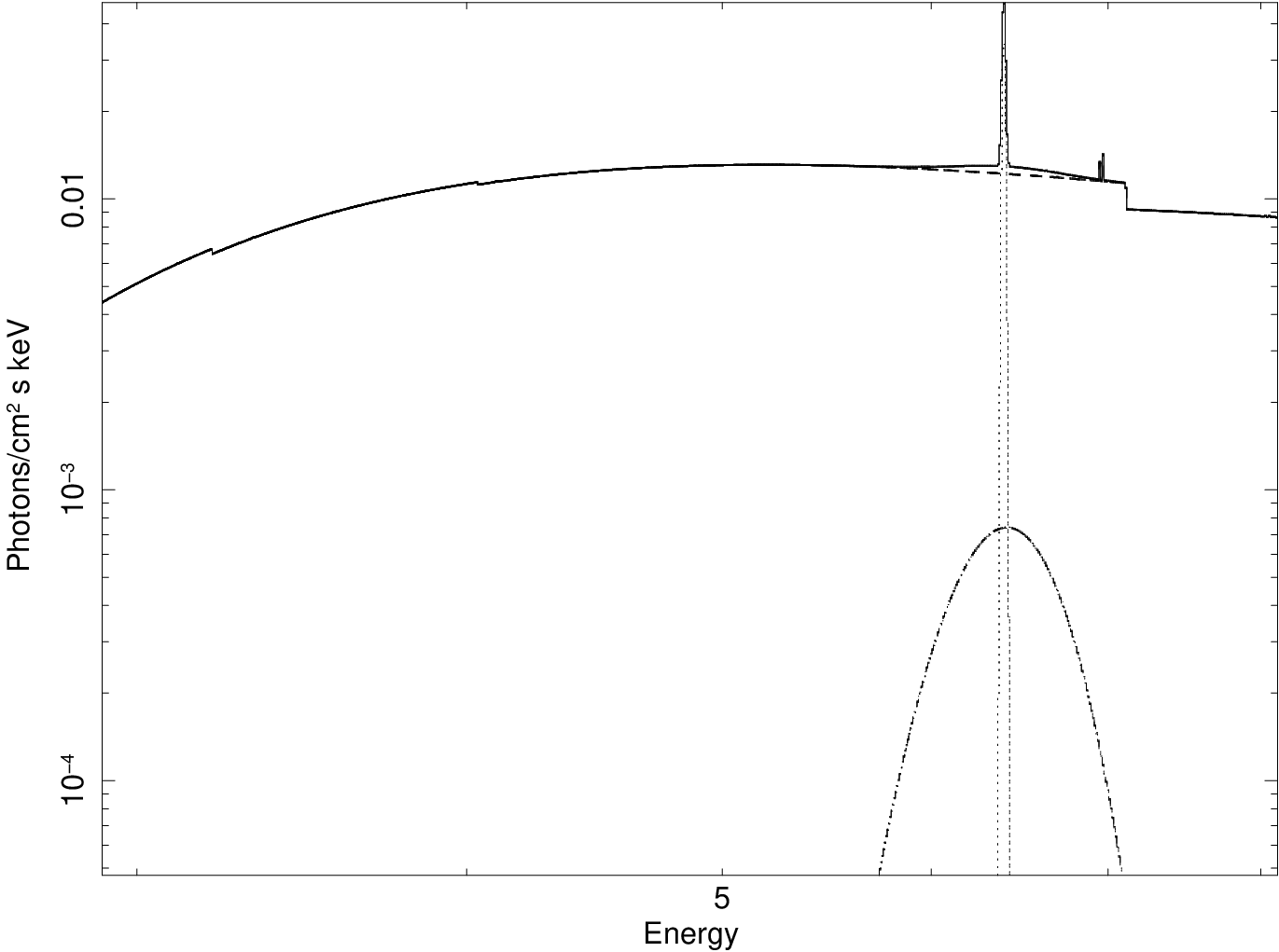}
\caption{Left: HETG spectra of the HMXB 4U1700$-$37 for $m=-1$ and $m=+1$ orders. Right: model used to fit the data including a narrow component and a broad component. The parameters of the broad component are taken from a non simultaneous {\it RXTE} observation of this source. Its presence though, is not required at all by the \chandra\ data.}
\label{fig:narrowbroad}
\end{figure*}

We emphasize that the present study deals specifically with the
\emph{narrow} \fe\ \ka\ line, for which the \chandra\ HETG instrument
is specifically well suited. A detailed analysis of the broad
 component would require simultaneous coverage by other broadband telescopes, like {\it
  RXTE}.  To illustrate how, we show in Fig. \ref{fig:narrowbroad} the HETG spectra of 4U1700$-$37 in the region under study (left). Here however, we have extended the spectral range to allow for the detection of the (putative) broad component. We have fitted the data with an absorbed powerlaw and an Fe line composed by a narrow and a broad component (right). For the broad component we have used the line parameters obtained from a non simultaneous {\it RXTE} observation of this source ($E=6.4$ keV, $FWHM=0.47$ keV, $EW=133$ eV, keeping the norm equal for both components). It turns out that the broad component is only required by {\it RXTE-PCA} data but not required at all by the \chandra\ data even though, in principle, the spectral range covered should be enough for this purpose, as can be clearly seen in Fig. \ref{fig:narrowbroad}. A similar effect and discussion can be seen in \cite{hanke09}. As the vast majority of our sample does not have
simultaneous \chandra - {\it RXTE} observations, no attempt
is made in the present study to involve the spectral continuum in
excess of the 4.96--7.74 keV (1.6 -- 2.5 \AA) band.

\begin{deluxetable*}{llccrrrc}
\tabletypesize{\scriptsize}
\tablecaption{Parameters for \fe\ \ka\ fluorescence line\label{tab:ka}}
\tablewidth{0pt}
\tablehead{
\colhead{Source} & \colhead{ObsID} & \colhead{$F_{1.6-2.5\rm \AA}$} & \colhead{$F(K\alpha)$} & \colhead{$EW(K\alpha)$} &
\colhead{$\lambda_{K\alpha}$} & \colhead{$\tau_{\rm edge}$} & \colhead{$N^{\rm equiv}_{\rm H}$} \\
 & &$\times 10^{-10}$ & $\times 10^{-4}$ & & & & $\times 10^{22}$ \\
  &    &  \colhead{(erg s$^{-1}$ cm$^{-2}$)} & \colhead{(photons s$^{-1}$ cm$^{-2}$)} & \colhead{(eV)} & \colhead{(\AA)} &   &  \colhead{(atoms cm$^{-2}$)}
}
\startdata
\cutinhead{HMXB}
OAO1657$-$415 & 1947 & 1.09$\pm$0.09 & 5.20$\pm$2.01 & 147.09$\pm$56.71 & 1.9366$\pm$0.0037 & 1.07$\pm$0.47 & 55.20$\pm$ 24.34 \\

Cen X-3       & 705  & 1.55$\pm$0.03 & 1.87$\pm$0.50 & 47.32$\pm$12.52 & 1.9425$\pm$0.0020 & 0.62$\pm$0.09 & 28.81$\pm$ 4.17 \\
              & 1943 & 23.14$\pm$2.08 & 10.69$\pm$1.37 & 17.43$\pm$2.29 & 1.9375$\pm$0.0011 & 0.18$\pm$0.03 & 11.13$\pm$ 0.92 \\

Cyg X-3       & 1456 & 10.36$\pm$2.49 & 22.00$\pm$2.64 & 65.28$\pm$7.88 & 1.9357$\pm$0.0016 & 0.66$\pm$0.06 & 28.77$\pm$ 2.60 \\
              & 425  & 29.17$\pm$3.50 & 14.78$\pm$4.35 & 15.64$\pm$4.61 & 1.9366$\pm$0.0029 & 0.34$\pm$0.05 & 9.82$\pm$ 0.19 \\
              & 426  & 26.65$\pm$0.16 & 18.78$\pm$7.06 & 21.48$\pm$8.07 & 1.9398$\pm$0.0057 & 0.36$\pm$0.05 & 25.43$\pm$ 2.25 \\
Cyg X-1       & 3815 & 27.13$\pm$2.98 & 16.80$\pm$1.82 & 19.07$\pm$2.06 & 1.9397$\pm$0.0018 & 0.03$\pm$0.01 & 1.80$\pm$ 0.73 \\
                    & 2415 &  17.65$\pm$1.68  & 12.43$\pm$1.72 & 20.82$\pm$2.05 & 1.9352$\pm$0.0092 & 0.07$\pm$0.03 & 3.31$\pm$1.25  \\
X1908$+$075   & 6336 & 0.45$\pm$0.35 & 1.23$\pm$0.50 & 81.02$\pm$3.20 & 1.9375$\pm$0.0038 & 0.34$\pm$0.09 & 15.31$\pm$ 4.03 \\
              & 5476 & 0.99$\pm$0.03 & 3.60$\pm$0.67 & 79.79$\pm$24.85 & 1.9389$\pm$0.0025 & 0.10$\pm$0.05 & 4.02$\pm$ 2.03 \\
              & 5477 & 0.34$\pm$0.18 & 0.69$\pm$0.33 & 58.06$\pm$26.60 & 1.9380$\pm$0.0037 & 0.67$\pm$0.30 & 29.45$\pm$ 13.13 \\
Vela X-1      & 1928 & 11.41$\pm$0.08 & 20.60$\pm$1.50 & 53.93$\pm$3.94 & 1.9375$\pm$0.0004 & 0.06$\pm$0.04 & 3.29$\pm$ 2.19 \\
              & 1927 & 9.38$\pm$0.14 & 34.00$\pm$1.73 & 107.04$\pm$5.50 & 1.9384$\pm$0.0003 & 0.38$\pm$0.05 & 18.55$\pm$ 1.98 \\
              & 1926\tablenotemark{(b)} & 0.08$\pm$0.07 & 1.94$\pm$0.22 & 931.51$\pm$112.76 & 1.9391$\pm$0.0010 & 0.85$\pm$0.38 & 34.78$\pm$ 15.51 \\
& 102\tablenotemark{(b)} & 0.09$\pm$0.08 & 1.50$\pm$0.39 & 625.63$\pm$166.80 & 1.9392$\pm$0.0029 & 0.70$\pm$0.35 & 28.49$\pm$ 14.20 \\
4U1700$-$37   & 657  & 3.63$\pm$0.04 & 8.70$\pm$0.81 & 70.19$\pm$6.45 & 1.9386$\pm$0.0006 & 0.30$\pm$0.05 & 15.03$\pm$ 2.52 \\
GX301$-$2     & 3433 & 7.52$\pm$0.05 & 32.00$\pm$1.14 & 113.54$\pm$4.08 & 1.9384$\pm$0.0002 & 0.48$\pm$0.03 & 25.15$\pm$ 1.80 \\
              & 2733 & 7.52$\pm$0.08 & 78.15$\pm$3.39 & 282.68$\pm$3.30 & 1.9388$\pm$0.0002 & 1.54$\pm$0.01 & 82.02$\pm$ 0.53 \\
              & 103  & 1.38$\pm$0.07 & 8.70$\pm$0.91 & 166.78$\pm$19.65 & 1.9392$\pm$0.0006 & 1.22$\pm$0.02 & 61.82$\pm$ 0.76 \\
LMC X-4       & 9571 & 0.31$\pm$0.02 & 0.83$\pm$0.22 & 73.71$\pm$21.95 & 1.9374$\pm$0.0054 & 0.23$\pm$0.15 & 9.09$\pm$ 5.00 \\
              & 9573 & 0.02$\pm$0.01 & 0.47$\pm$0.19 & 890.00$\pm$267.00 & 1.9422$\pm$0.0056 & 0.74$\pm$0.35 & 24.33$\pm$ 12.00 \\
              & 9574\tablenotemark{(b)} & 0.07$\pm$0.01 & 0.45$\pm$0.25 & 243.03$\pm$37.74 & 1.9409$\pm$0.0035 & 1.43$\pm$0.40 & 58.42$\pm$ 10.00 \\
$\gamma$ Cas\tablenotemark{(a)}  & 1985 & 0.52$\pm$0.01 & 0.71$\pm$0.27 & 39.74$\pm$15.04 & 1.9368$\pm$0.0001 & 0.57$\pm$0.23 & 22.76$\pm$1.06 \\
\cutinhead{LMXB}
4U1822$-$37   & 671 & 2.32$\pm$0.53 & 2.94$\pm$0.69 & 36.61$\pm$8.52 & 1.9379$\pm$0.0015 & 0.31$\pm$0.08  & 15.52$\pm$3.72  \\
GX1$+$4       & 2710 & 0.66$\pm$0.01 & 1.64$\pm$0.38 & 72.81$\pm$16.96 & 1.9376$\pm$0.0013 & 0.26$\pm$0.19 & 12.51$\pm$8.62  \\
Her X-1       & 2749\tablenotemark{(b)} & 0.20$\pm$0.03 & 3.00$\pm$0.40 & 512.88$\pm$68.98& 1.9372$\pm$0.0037 & 0.38$\pm$0.37 & 15.26$\pm$15.00 \\
              & 3821 & 0.74$\pm$0.16 & 3.49$\pm$0.74 & 141.02$\pm$29.76& 1.9375$\pm$0.0015 & 0.48$\pm$0.22 & 24.32$\pm$10.89 \\
              & 3822 & 0.94$\pm$0.63 & 1.25$\pm$0.83 & 162.56$\pm$107.94 & 1.9386$\pm$0.0013 & 0.70$\pm$0.21 & 33.84$\pm$10.36 \\
              & 4585 & 2.15$\pm$1.15 & 3.74$\pm$1.41 & 51.58$\pm$19.46 & 1.9399$\pm$0.0031 & 0.64$\pm$0.15 & 30.79$\pm$7.29 \\
              & 6149 & 2.32$\pm$0.53 & 5.48$\pm$1.26 & 69.05$\pm$15.88 & 1.9395$\pm$0.0002 & 0.45$\pm$0.12 & 22.42$\pm$6.09 \\
              & 6150 & 0.75$\pm$0.20 & 3.18$\pm$0.85 & 128.59$\pm$34.29 & 1.9397$\pm$0.0018 & 0.74$\pm$0.30 & 34.58$\pm$14.28\\ 
Cir X-1   & 706     & 66.31$\pm$4.64 & 13.94$\pm$ 0.03 & 6.07$\pm$1.50 & 1.9392$\pm$0.0018 & 0.14$\pm$0.02 & 7.47$\pm$1.23\\
               & 1700   & 36.59$\pm$0.46 & 17.68$\pm$0.03 & 14.53$\pm$1.21 & 1.9388$\pm$0.0019 & 0.15$\pm$0.03 & 9.42$\pm$1.35
\enddata
\tablecomments{Only sources and ObsIDs. with positive detections have been included}
\tablenotetext{(a)}{Status as a HMXB still unclear. Not included in the correlations.}
\tablenotetext{(b)}{Eclipse data}
\end{deluxetable*}


\begin{deluxetable*}{lrrrr}
\tabletypesize{\scriptsize}
\tablecaption{Parameters for \fe\ \kb\ fluorescence line \label{tab:kb}}
\tablewidth{0pt}
\tablehead{
\colhead{Source} & \colhead{ObsID}  & \colhead{$F(K\beta)(\times 10^{-4})$} & \colhead{$EW(K\beta)$} &
\colhead{$\lambda_{K\beta}$}  \\
  &    &  \colhead{(photons s$^{-1}$ cm$^{-2}$)} & \colhead{(eV)} & \colhead{(\AA)} 
}
\startdata
OAO1657$-$415 & 1947 & 0.73$\pm$0.73 & 86.94$\pm$33.52 & 1.7500$\pm$0.0100 \\ 
Cen X-3 & 705 & 0.25$\pm$0.35 & 5.90$\pm$2.50 & 1.7577$\pm$0.0100 \\
Cyg X-1 & 3815 & 5.37$\pm$2.48 & 6.23$\pm$0.67 & 1.7554$\pm$0.0069 \\  
X1908$+$075 & 6336 & 1.08$\pm$1.48 & 77.72$\pm$3.07 & 1.7464$\pm$0.0057 \\ 
            & 5476 & 1.20$\pm$1.48 & 41.02$\pm$22.70 & 1.7453$\pm$0.0051 \\
            & 5477 & 0.24$\pm$2.18 & 17.92$\pm$8.21 & 1.7471$\pm$0.0076 \\
Vela X-1    & 1928 & 4.00$\pm$2.13 & 11.30$\pm$0.83 & 1.7597$\pm$0.0034 \\
            & 1927 & 3.90$\pm$2.03 & 11.82$\pm$0.61 & 1.7556$\pm$0.0037 \\
            & 1926 & 0.22$\pm$0.19 & 95.88$\pm$11.61 & 1.7553$\pm$0.0099 \\
4U1700$-$37 & 657  & 1.40$\pm$1.93 & 11.06$\pm$1.02 & 1.7595$\pm$0.0057 \\
GX301$-$2   & 3433 & 3.50$\pm$1.17 & 11.13$\pm$0.40 & 1.7550$\pm$0.0024 \\
            & 2733 & 11.8$\pm$2.45 & 41.10$\pm$0.48 & 1.7569$\pm$0.0014 \\
            & 103  & 1.32$\pm$0.75 & 16.07$\pm$1.89 & 1.7578$\pm$0.0038 \\
LMC X-4     & 9571 & 0.09$\pm$0.02     & 8.39$\pm$1.72 & 1.7400 (frozen) \\
                   & 9574 & 0.19$\pm$ 0.03    & 72.43$\pm$14.49 & 1.7528 (frozen) \\
Her X-1      & 4585 & 1.45$\pm$  0.05   & 18.51$\pm$2.31 & 1.7501 (frozen) 
\enddata
\end{deluxetable*}

\section{Results}

The results are presented in Tables \ref{tab:ka}, \ref{tab:kb} and
\ref{tab:compt}. They allow for several immediate conclusions to be
drawn.  Fe fluorescence emission seems to be quite ubiquitous in
HMXB. All 10 HMXB analyzed showed fluorescence emission. Those sources
with several ObsIDs available show that this emission is highly
variable. A particularly striking case is that of Cyg X-1 where only
two out of nine observations contained only very weak detections. 

In contrast, Fe fluorescence emission seems to be very rare amongst
LMXB. Only four LMXB sources out of the 31 analyzed, showed
fluorescence line emission. Apart from that, there are no obvious
differences in the detected emission levels in LMXB compared with the
ones in HMXB.  To observe Fe K line fluorescence in $\gtrsim 90\%$ of
the HMXB observations, but only in $\gtrsim 10\%$ in LMXBs is in
contrast with the finding of \cite{gott95} and \cite{asai00}.

The fluorescence Fe \ka\ line is centered at $\lambda_{\rm Fe
  K\alpha}=1.9387\pm 0.0016$ \AA, on average, and we do not see
significant shifts to higher ionization states.  This is equivalent to
an energy range from 6.390 - 6.400 keV, consistent with the two
components K$\alpha_{1}$ and K$\alpha_{2}$ from ion states below
Fe\textsc{ x} \citep{house69}.  These two components are not resolved
and the line appears narrow in all cases with $FWHM\lesssim 0.005$
\AA, while the lines in \cite{gott95} and \cite{asai00} are of the
order of $\sim 0.5-1$ keV.  Last, but not least, a Compton shoulder is
detected in the only hypergiant source in the sample,
GX301-2~\citep{watanabe03} and also in X1908+075, in the latter case,
for the first time.

\section{Curve of Growth}

\begin{figure}
\includegraphics[angle=0,width=\columnwidth]{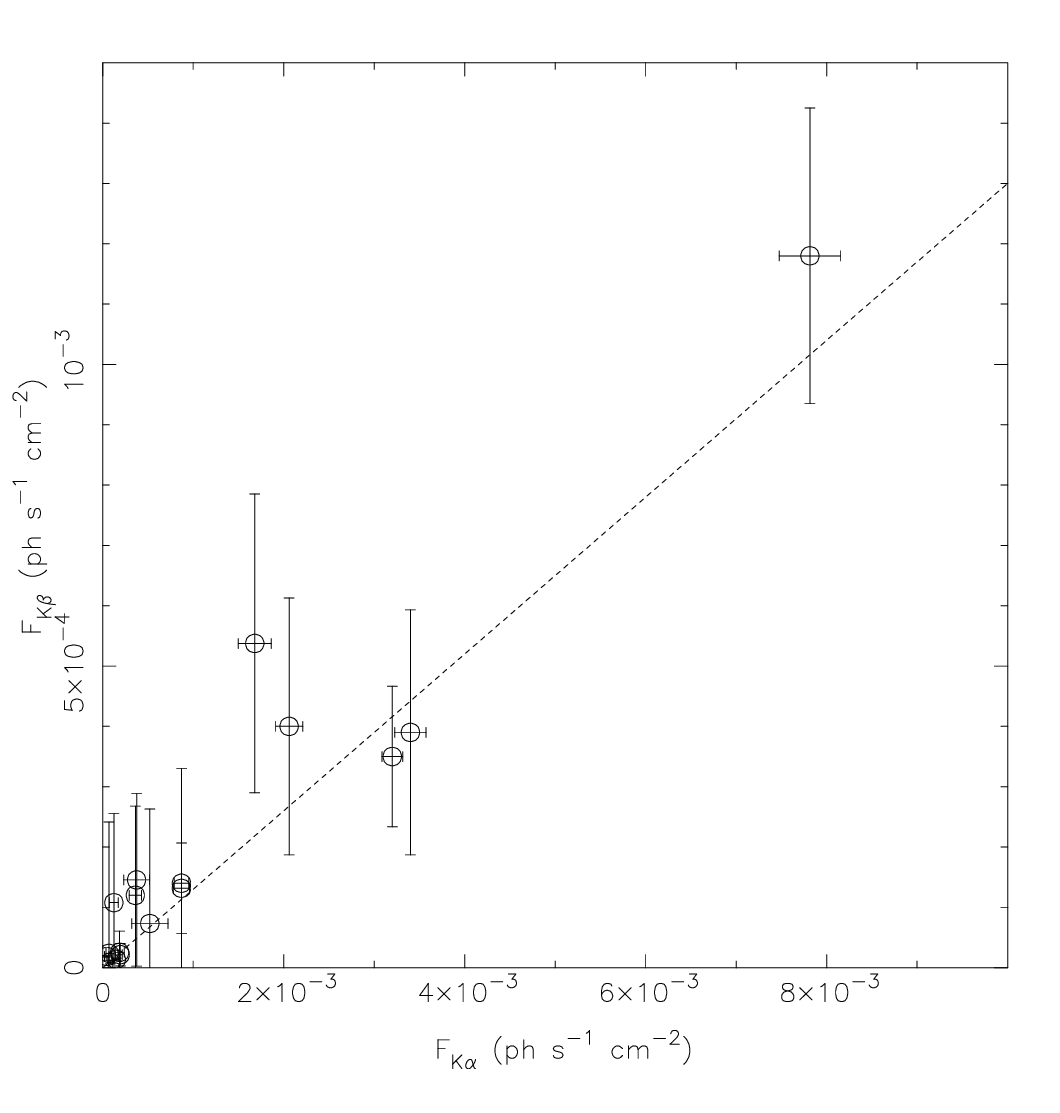}
\caption{The measured fluxes of the \fe\ \kb\ line versus the \fe\ \ka\ line where both lines could be detected   simultaneously. The dashed line represents the theoretical prediction \citep{palmeri03}.}
\label{fig:fafb}
\end{figure}

\begin{figure*}
\includegraphics[angle=0,width=\columnwidth]{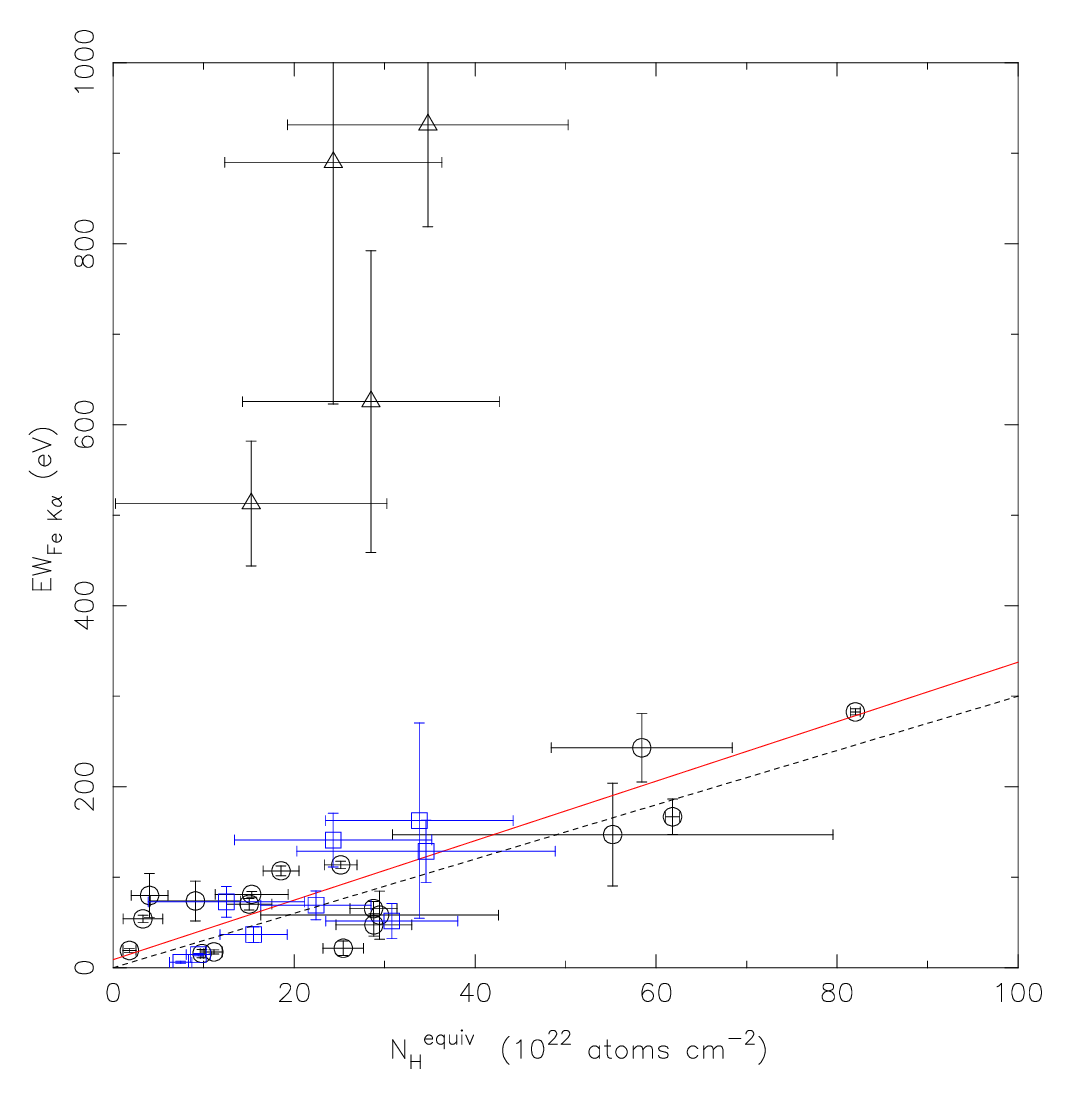}
\includegraphics[angle=0,width=\columnwidth]{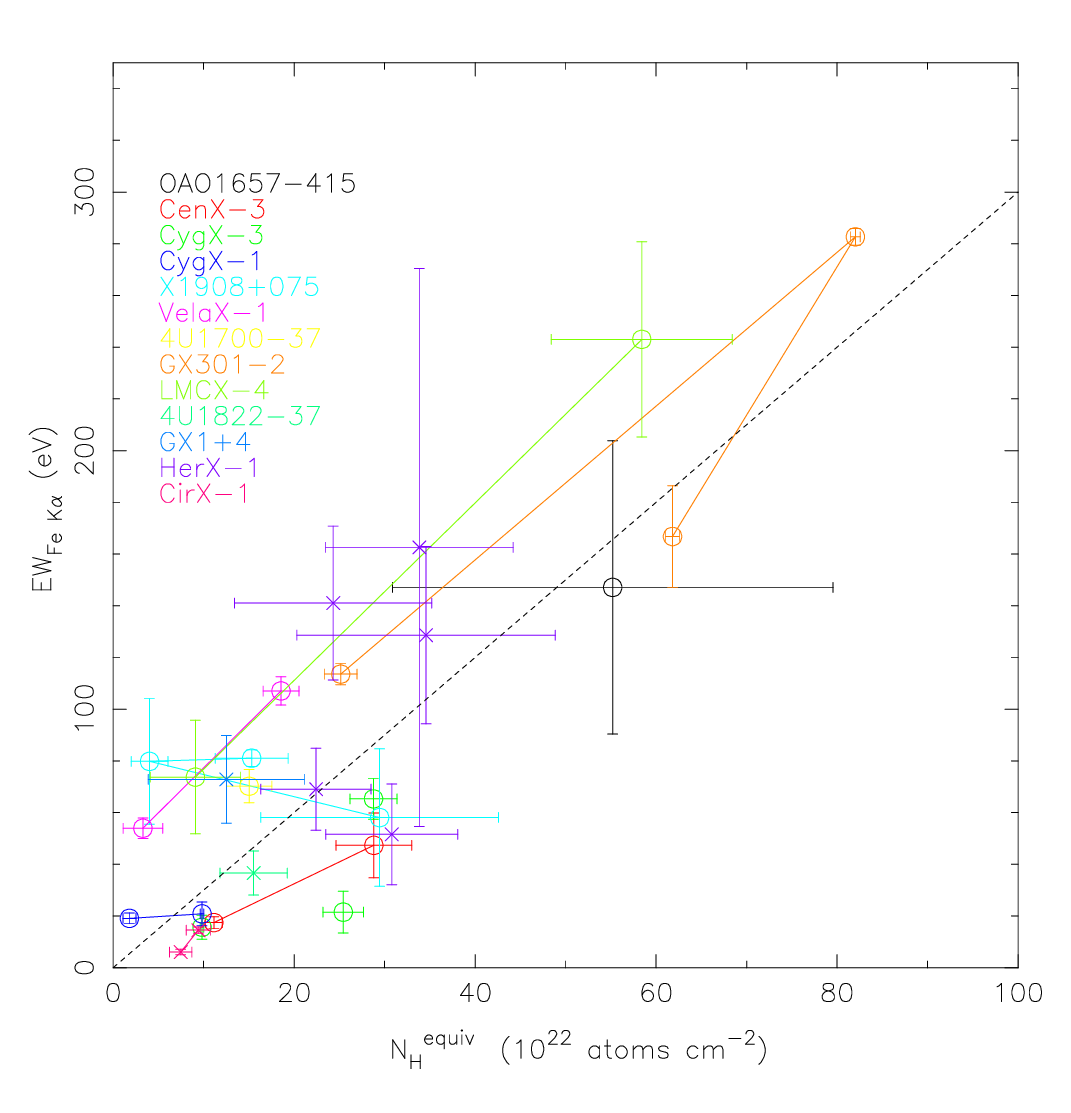}
\caption{Left panel: The curve of growth. The $EW$ of the Fe line grows with the
  column density of the reprocessing material (continuous red line) as predicted
  by the theoretical model of \cite{kallman04} for spherical geometry
  (dashed line). Black circles represent HMXB data and blue squares
  represent LMXB. Eclipse data (triangles, upper left corner) do not
  follow the main trend. Right panel: the several ObsIDs for the same source have been linked by a line, showing that, with few exceptions, the sources follow the relationship individually.}
\label{fig:cog}
\end{figure*}

One of the goals of the this study is to establish an empirical
relationship between $EW$ of the \fe\ \ka\ fluorescence line and the
column density of the reprocessing material and determine its Curve of
Growth.  In Fig. \ref{fig:fafb} we plot the measured fluxes of
\fe\ \kb\ versus \ka\ (from data in Tables \ref{tab:ka} and
\ref{tab:kb}) for those sources where both lines could be reliably
measured simultaneously. As can be seen, the data do not deviate
significantly from the theoretical value $F(K\beta)/F(K\alpha)\sim
0.13-0.14$ \citep{palmeri03}. Therefore the sample does not show
strong evidence of excess resonant Auger destruction and the derived
$N_{\rm H}$ will be reliable. The exception
is X1908$+$07 for which this ratio seems to be larger.  In Fig
\ref{fig:cog} we plot the curve of growth of the $EW(K \alpha)$ with
the H equivalent absorption column of the reprocessing material.  As
can be seen, the {\it EW}s of the Fe \ka\ lines scale linearly with $N_{\rm
  H}$. The least squares best fit gives

\begin{equation}
\label{eq:ewnh}
EW_{Fe K\alpha}({\rm eV})=(3.29\pm0.05)N^{22}_{\rm H} 
\end{equation}

where $N^{22}_{\rm H} $ is the equivalent H column of the reprocessing
material in units of $10^{22}$ atoms cm$^{-2}$. This relationship is
plotted as a continuous red line in Fig \ref{fig:cog}. The degree of
correlation is very high (Pearson correlation coefficient of
$r=0.95$). We have also plotted the theoretical prediction given by
\cite{kallman04}, Eq.5, namely $EW(\fe K\alpha)\simeq 3N^{22}_{\rm H}$ [eV]
(black dashed line). The agreement is excellent even though there is a
large scatter in the data.  This theoretical equation has been
computed for spherical geometry and solar abundance. The available
data, therefore, are consistent with a spherical distribution of the
Fe fluorescence emission zone around the X-ray source. This implies a
scenario where the compact object is deeply embedded within the
stellar wind of the companion star, as is the likely case in most
HMXBs. It is very important to stress that to the most extent, several
individual observations of the same object follow this relationship (Fig.\ref{fig:cog}, right panel).
The supergiant HMXB X1908$+$07 is an exception and, in fact, shows an
\emph{anti}correlation. Likewise, no saturation
effects are observed within the range of columns studied. Our data in
Fig. \ref{fig:cog} are also consistent with the lines I and II of
Fig. 4 in \cite{inoue85} where the reprocessing matter is either
spherically distributed around the X-ray source or located between the
X-ray source and the observer. Since the compact object is deeply
embedded into the wind of the donor, the actual situation is a
combination of both. There are points that deviate significantly from
this trend (shown as triangles). These points correspond to eclipse
data of Vela X-1 and LMC X-4 (ObsId 102, 1926 and 9573 respectively).
This can be explained by the computation of the $EW$ 
\citep{kallman04}, which states:

\begin{equation}
EW=\frac{N\omega_{K}y_{Fe}x_{l}}{f_{\epsilon K}}\int_{\epsilon_{\rm Th}}^{\infty}f_{\epsilon}\sigma_{\rm K}(\epsilon)\frac{d\epsilon}{\epsilon}
\end{equation}

where $N$ is the radial column density of the reprocessing material, $\omega_{K}$ is the fluorescence yield, $y_{Fe}$ is the Fe elemental abundance, $\sigma_{\rm K}(\epsilon)$ is the K shell photoionization cross section and 
$f_{\epsilon}$ is the local ionizing continuum. 
The intensity of the line is normalized to the local continuum,
which is strongly suppressed during eclipse. The illuminating source
is not observed directly either. The line emitting gas is exposed to the full continuum from the compact object and we see the line emitting gas directly. However, we observe the continuum only via scattering.  As a consequence, in eclipse, the
very large $EW$ does not correlate well with $N_{\rm H}$.  We
therefore do not use eclipse points for the subsequent analysis.



Instead of relating the {\it EW} and the column density, it might be useful
to determine the relationship between the {\it EW} and the optical depth at
the K edge. From our data we obtain

\begin{equation}
\label{eq:tauew}
\tau_{K~edge}=(0.568\pm0.005)\left(\frac{EW_{Fe K\alpha}}{100 \rm ~eV}\right)
\end{equation}

with a Pearson correlation coefficient of $r=0.98$.  This equation
shows that the reprocessing material reaches optical depth unity for
emissions of $EW\sim 175$ eV. Combining Equations \ref{eq:ewnh} and
\ref{eq:tauew} we recover the ISM cross section at the K edge, namely
$\sigma_{\rm Fe}=1.8\times 10^{-24}$ cm$^{-2}$ (\cite{wilms00}) as
might be expected. It has been shown in \cite{waldron98} that
the continuum cross section of the wind in OB stars $\sigma_{\rm W}$
is always less than the interstellar value $\sigma_{\rm ISM}$. This is
due to the intense UV radiation emitted by the star which increases
the ionization state with respect the ISM. This difference is large at
low energies while for energies above 1.5 keV ($\lambda \leq 8.3$ \AA)
we have $\sigma_{\rm W} \approx \sigma_{\rm ISM}$. The fact that we
get the ISM value of \cite{wilms00} is an independent check of the
above statement and consistent with the compact object deeply embedded
into the wind of the donor.

In conclusion, we find that the curve of growth is consistent with a
spherical distribution of reprocessing material around the X-ray
source and follows the theoretical prediction of \cite{kallman04}
(Eq. 5), namely $EW(\fe K\alpha)\simeq 3N^{22}_{\rm H}$ [eV]. This is
further supported by the fact that the Fe lines are very narrow and in
most cases not resolved by \chandra. This means that the material is
not rotating at high speeds as would be the case in accretion disks.  The
few LMXBs follow the same trend as the HMXB. In LMXBs, though, the
situation must be different to that in HMXB where the neutron stars
are deeply embedded into the stellar winds of their massive
companions. To imply a spherical geometry in the LMXB cases is less
straight forward but it means that through some mechanism the
fluorescing material has lost all 'memory' of the donor.  
The eclipse data
seem to follow a different relationship with enormous $EW$ which do
not match the corresponding large $N_{\rm H}$ deduced from the
equation in point 1.


\section{On the X-ray Baldwin Effect in XRBs}

\begin{figure}
\includegraphics[angle=0,width=\columnwidth]{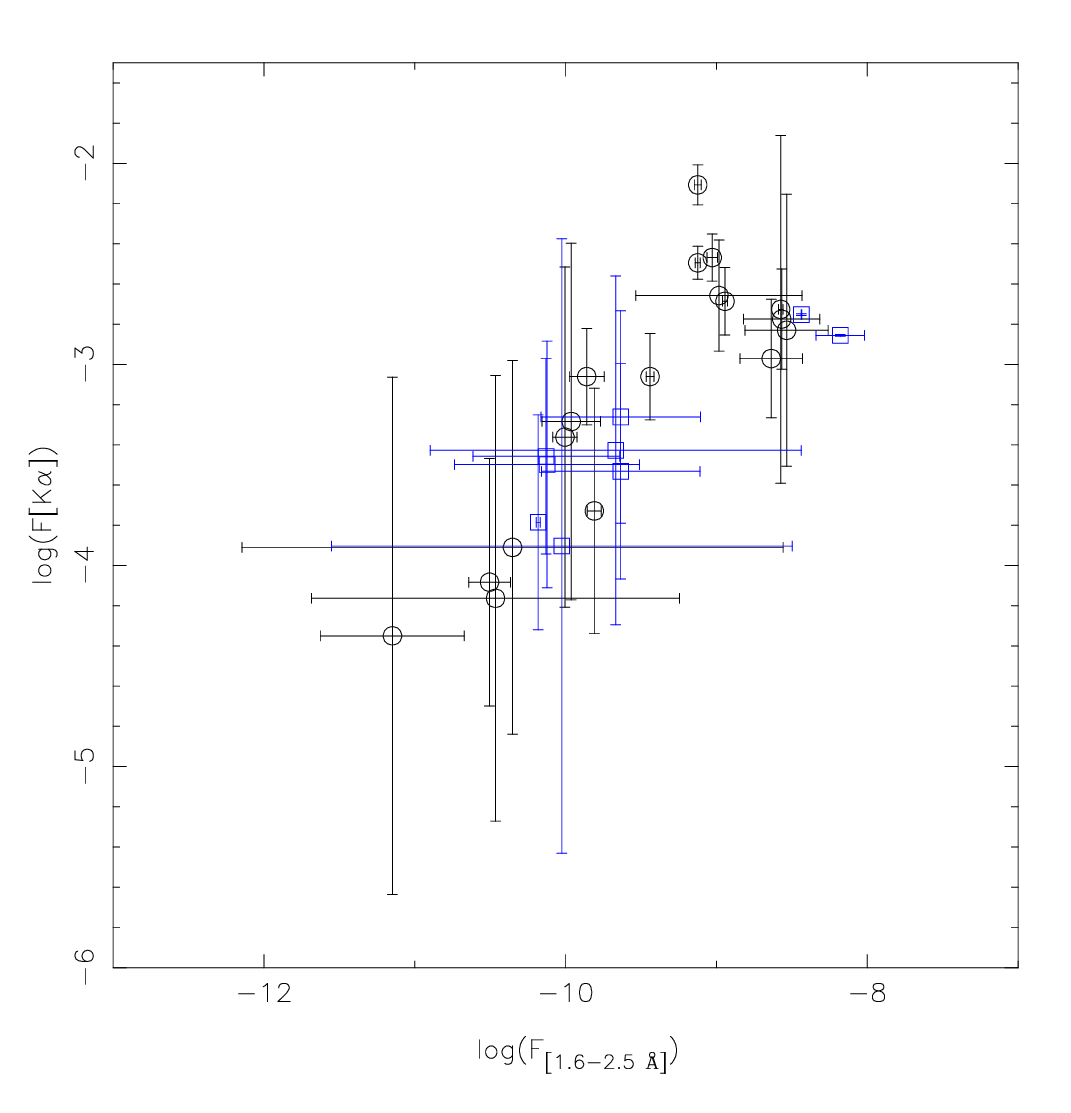}
\caption{Log-Log plot of the line flux vs the continuum flux. A good
  correlation exists ($F_{line}\propto F_{cont}^{0.71};
  r^{2}=0.75$).The reprocessed fluorescence photons variations track
  closely the continuum emitted by the X-ray source. This means that
  the reprocessing region is very close to the X-ray continuum
  source. Black circles represent HMXB and blue squares LMXB. }
\label{fig:fcont_fline}
\end{figure}

In Fig. \ref{fig:fcont_fline} we show the correlation between the
unabsorbed continuum flux in the 1.6 - 2.5 \AA\ band, the most
effective source for Fe K fluorescence , and the Fe K$\alpha$ line
flux.  As can be seen , the variations of the reprocessed fluorescence
photons track closely the variations of the continuum emitted by the
illuminating X-ray source. This means that the reprocessing region
must be very close to the X-ray source. 

\begin{figure}
\epsscale{1.0}
\includegraphics[angle=0,width=\columnwidth]{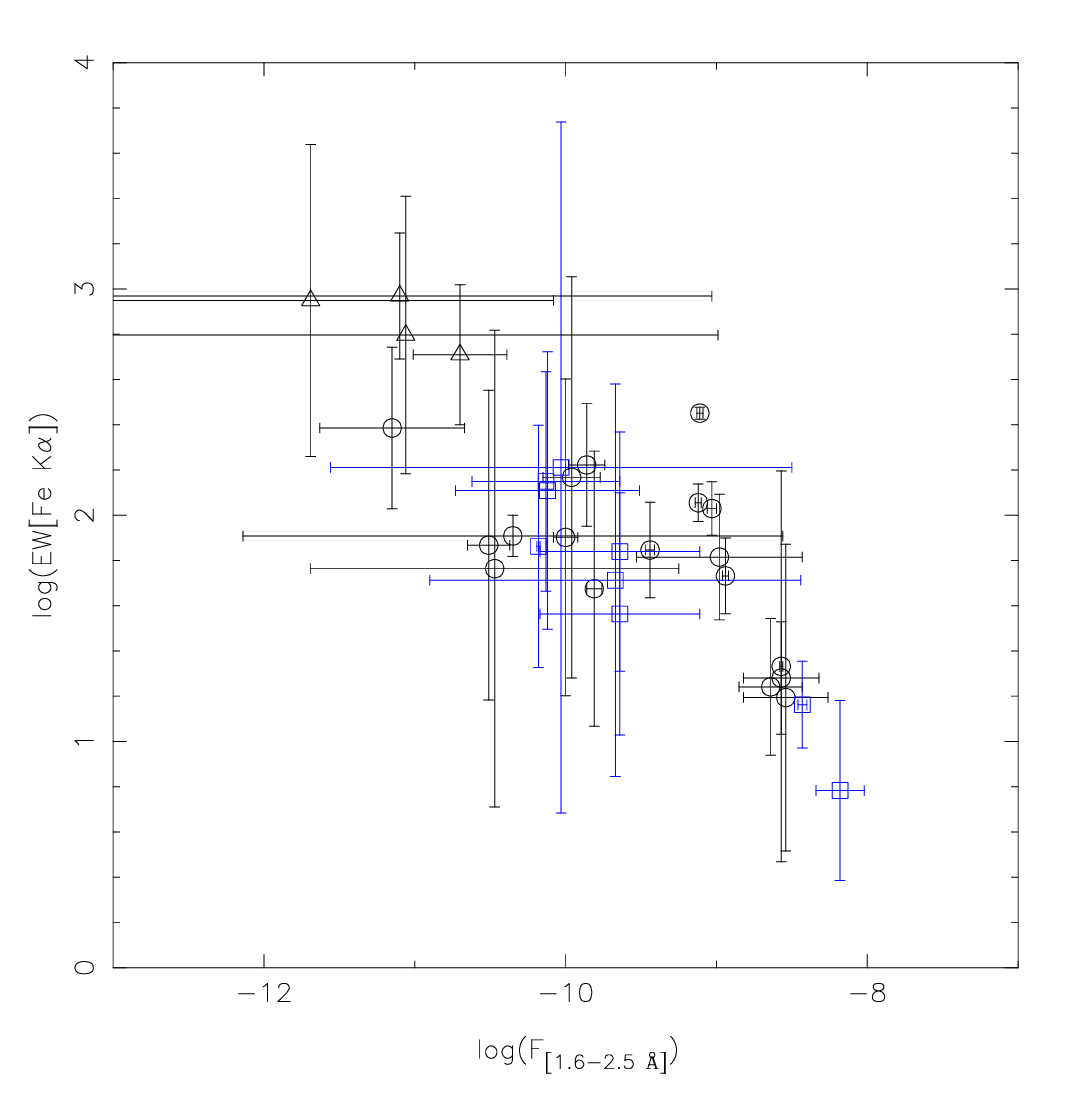}
\caption{Log-log plot of the EW of the Fe line vs the unabsorbed flux
  of the continuum. A clear anticorrelation can be seen. The larger
  the $F_{X}$ of the central source, the smaller the strength of the
  line. Black circles represent HMXB, blue squares LMXB and triangles
  the eclipse data. $EW_{line}\propto F_{cont}^{-0.437}; r^{2}=0.80$)}
\label{fig:baldwin_f}
\end{figure}

\begin{figure}
\epsscale{1.0}
\includegraphics[angle=0,width=\columnwidth]{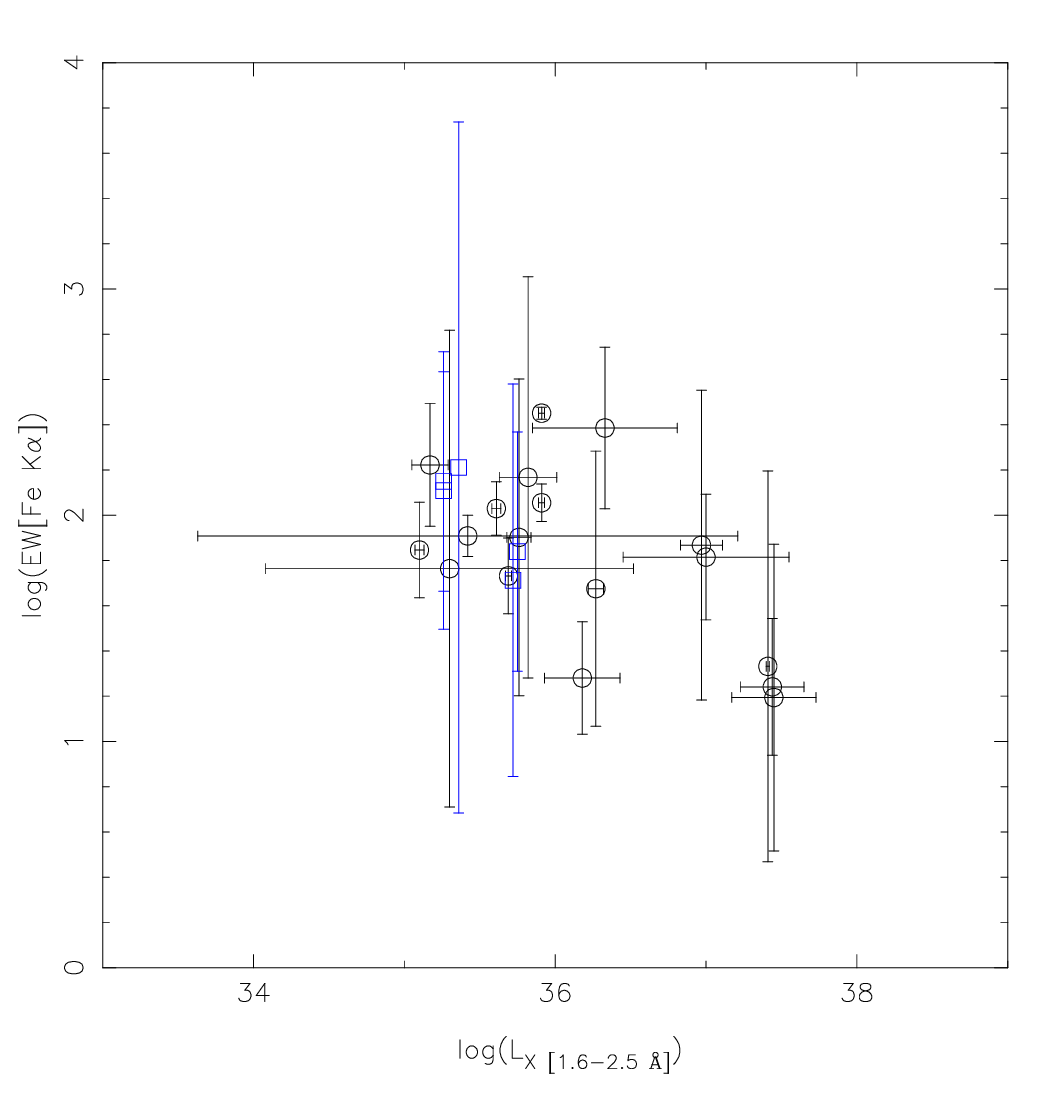}
\caption{Baldwin effect for HMXB. Only sources with known distances
  have been included. A anticorrelation can be seen although the
  correlation coefficient worsens with respect to that of
  Fig. \ref{fig:baldwin_f}. This is expected as the larger the $L_{X}$
  of the central source, the more ionized is the Fe of the
  reprocessing region and, therefore, the smaller is the $EW$ of the
  fluorescence line. Eclipse data have not been
  plotted. $EW_{line}\propto L_{X}^{-0.289}; r^{2}=0.62$)}
\label{fig:baldwin_l}
\end{figure}

While we observe a correlation of line flux with continuum flux, we
also observe an \emph{anti}correlation of $EW$ with the same continuum
flux (Fig. \ref{fig:baldwin_f}).  This trend is also visible in the
\emph{EXOSAT} data of \cite{gott95} (Fig. 3). Such an anticorrelation
has been shown to exist for AGNs exhibiting Fe K$\alpha$ fluorescence
line (v.g. \cite{iwasawa93}, \cite{jiang06}, \cite{mattson07}).  This
phenomenon is called the 'X-ray Baldwin effect' following the
discovery of a decrease in the EW of the C\textsc{iv} line with
increasing UV luminosity in AGNs by \cite{baldwin77}. Note that in all
data sets there is a considerable scatter present. Then we compute the
intrinsic $L_{X}$ with the corresponding distance taken from the
literature.  Even though the trend is still visible, the correlation
begins to break down. This can be due to a number of reasons, which
may relate to issues such as that the distances for some systems are
poorly known, or, as shown recently by \cite{dunn08} for GX 339-4, the
X-ray Baldwin is effective in the soft state but not in the hard
state. Such issues can also contribute to the scatter in
Fig. \ref{fig:baldwin_l}.  These issues not withstanding, the
\chandra\ data show, for the first time, that such a correlation seems
to exist for X-ray binaries in general.

An immediate interpretation could be ( \cite{nayakshin00a,
  nayakshin00b}) that upon increase of the continuum X-ray flux from
the central source, the surrounding reprocessing material converts
from cold, neutral, to progressively ionized with the concurrent
decrease in the EW of the fluorescence Fe K$\alpha$ line.

\section{Asymmetric lines}

\begin{deluxetable*}{lrrrrrrr}
\tabletypesize{\scriptsize}
\tablecaption{Parameters for \fe\ \ka\ Compton shoulder\label{tab:compt}}
\tablewidth{0pt}
\tablehead{
\colhead{Source} & \colhead{ObsID}  & \colhead{$F_{\rm C}(\times 10^{-4})$} & \colhead{$EW_{\rm C}$} &
\colhead{$\lambda_{\rm C}$}  \\
  &    &  \colhead{(photons s$^{-1}$ cm$^{-2}$)} & \colhead{(eV)} & \colhead{(\AA)} 
}
\startdata
X1908$+$075 & 5476 & 1.74$\pm$0.64 & 75.91$\pm$27.92 & 1.9550$\pm$0.0031 \\ 
GX301$-$2   & 2733 & 31.23$\pm$4.68 & 178.35$\pm$26.75 & 1.9626$\pm$0.0037 \\
            & 3433 & 7.41$\pm$1.27 & 30.18$\pm$5.17 & 1.9559$\pm$0.0027 \\
\enddata
\end{deluxetable*}

\subsection{Compton Shoulders}

\begin{figure}
\includegraphics[angle=-90,width=0.49\columnwidth]{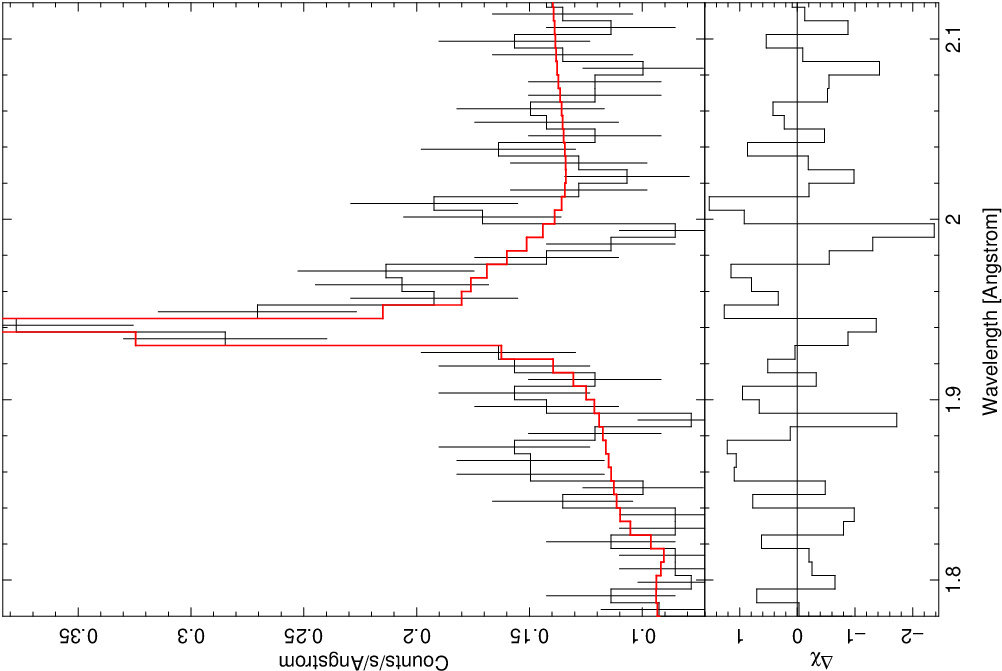}
\includegraphics[angle=-90,width=0.49\columnwidth]{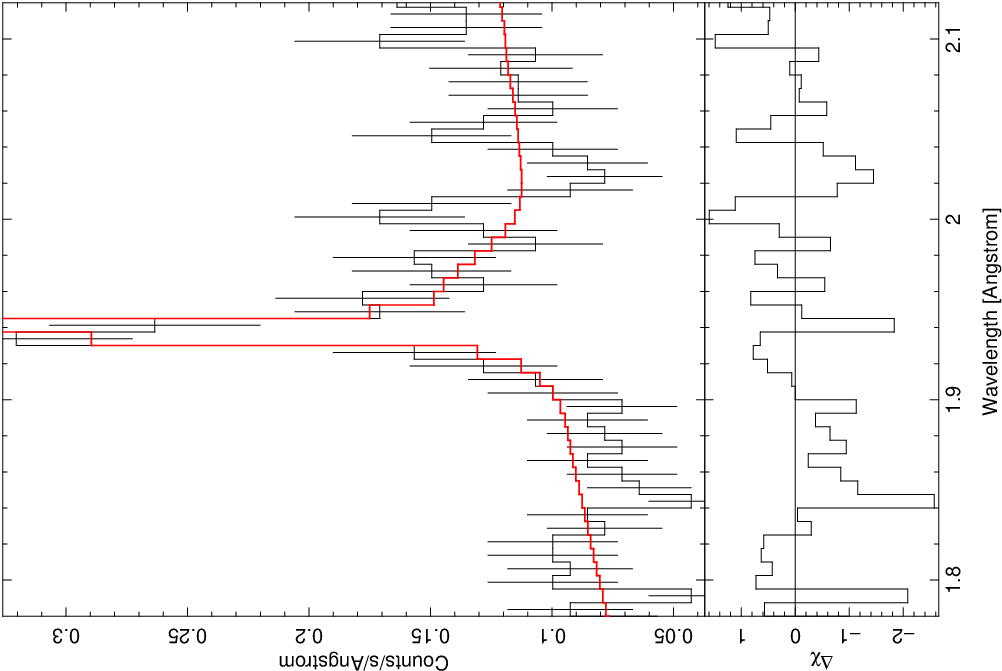}
\caption{Compton shoulder seen in X1908$+$075 (ObsId 5476) when the
  $N_{\rm H}$ is low for $m=-1$ (left) and $m=+1$ (right) orders
  respectively. The shoulder has been modeled with a simple gaussian
  and both orders have been fitted together. Parameters are given in
  Table \ref{tab:compt}. }
\label{fig:comptonX1908}
\end{figure}

Within the sample, some sources show asymmetric profiles. We detect a
significant shoulder in the supergiant binary X1908$+$075 during ObsId 5476.
In Fig. \ref{fig:comptonX1908} (HEG orders $m=\pm 1$), the
\fe\ \ka\ line shows an extension redward of the main line of $\Delta
\lambda\sim 0.06$ \AA\ up to $\sim 2$ \AA. The most natural
interpretation for this asymmetry is a Compton shoulder. Such features
have been resolved with \chandra\ high resolution capabilities for a
small number of extragalactic sources \citep{kaspi02},
\citep{bianchi02} and for GX 301-2~\citep{watanabe03}.  The high
energy photons produced in the fluorescence reprocessing region must
still traverse the circumsource and circumstellar material to reach
the interstellar medium. If this material is Compton thick, the
fluorescence photons have a probability of being Compton scattered off
electrons with the concurrent decrease in their energy. In
Fig. \ref{fig:comptonX1908} it is clear that primary K$\alpha$ photons
are downscattered down to $\sim 2$ \AA, i.e. around 0.06
\AA\footnote{although the center-to-center difference in the modeling
  gaussians is less than that: $1.955-1.937\approx 0.02$}. This is
exactly the maximum shift $\Delta
E_{max}=2E_{0}^{2}/(m_{e}c^{2}+2E_{0})$ with $E_{0}=6.40$ keV (=1.94
\AA), for back scattered photons ($\theta=180^{\rm o}$).  Therefore,
the shoulder is formed through single Compton scattering of the iron
K$\alpha$ photons. The apparent concavity of the shoulder is due to
the larger probability of forward and backward scatterings which
produce minimum and maximum energy shift respectively.

The flux ratio of the shoulder to the primary line is (see Tables
\ref{tab:ka} and \ref{tab:compt}) is $\sim 0.48$.  This is somewhat
higher than the value predicted by \cite{matt02}.  Curiously, the
shoulder is not observed during ObsIDs 5477 (high $N_{\rm H}$) and
6336 (medium $N_{\rm H}$).  As has been shown by \cite{watanabe03} and
\cite{matt02}, too high column densities tend to smear the shoulder as
second order scatterings become to be important.

\subsection{Asymmetric Line Wings}

\begin{figure}
\includegraphics[angle=-90,width=0.49\columnwidth]{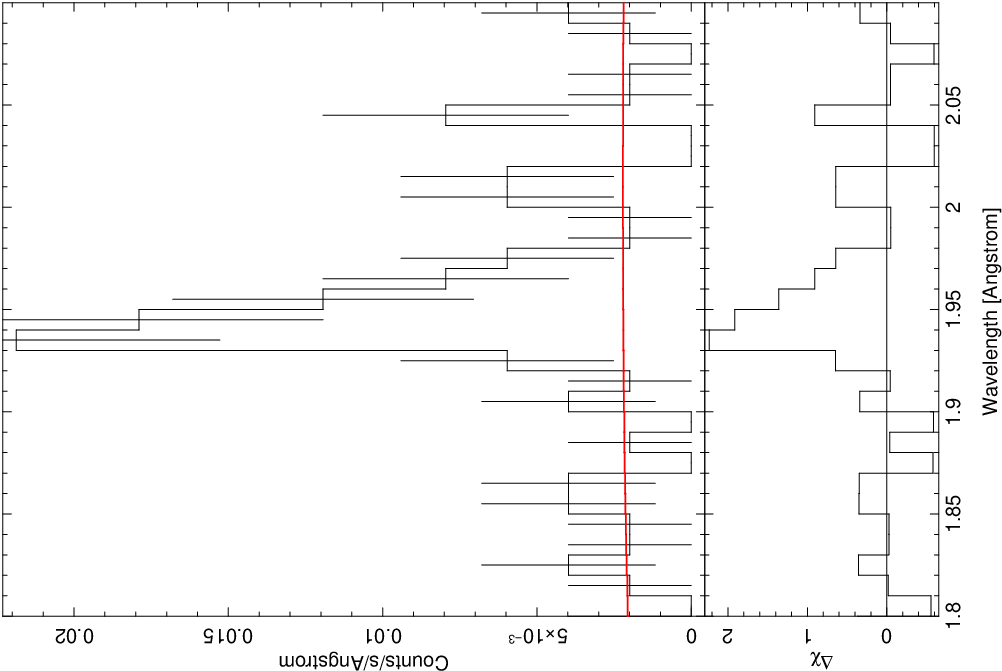}
\includegraphics[angle=-90,width=0.49\columnwidth]{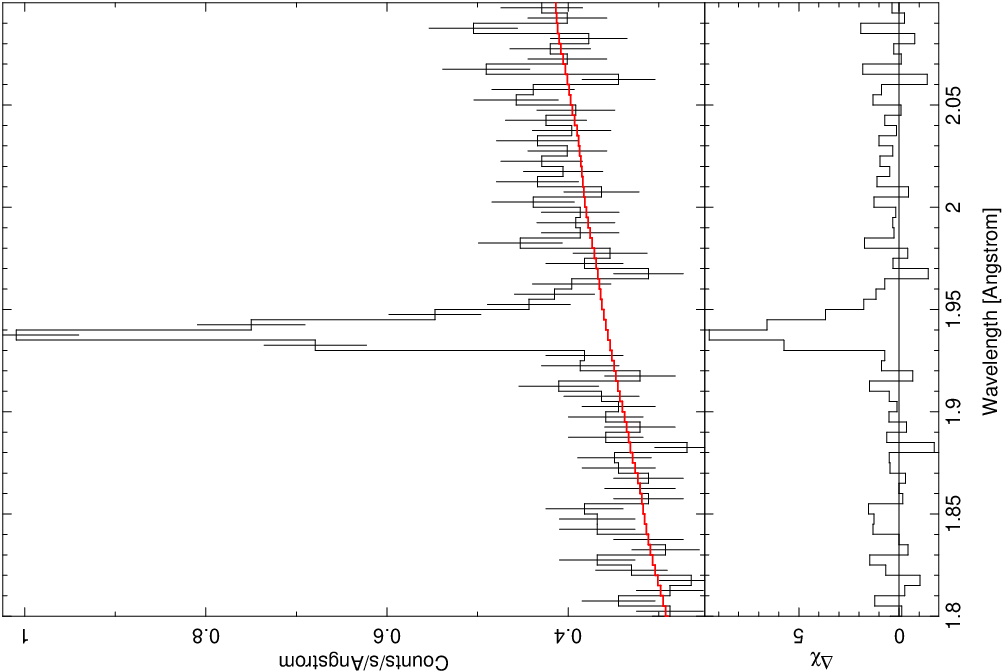}
\caption{Asymmetric profiles seen in LMC X-4 ObsId 9573 and
  4U1700$-$37 ObsId 657. The local continuum has been modeled with a
  powerlaw. No gaussians have been added to model the emission line so
  that the wing is clearly visible in the residuals.}
\label{fig:asym}
\end{figure}

Instead of the clear shoulder displayed in the previous section, some
sources present asymmetric wings which include a sharp blue side and
an extended, progressive red decline (see
Fig. \ref{fig:asym}\footnote{From the analysis point of view, the
  difference between Compton shoulder and asymmetric profiles is that
  the former require two gaussians to reduce the residuals while the
  latter only needed one.}).

In order to produce Compton shoulders, the scattering gas must be very
cold, i.e. less than $\sim 10^{5}$ K. However, anytime the gas is
cooler than $\sim E_{K\alpha}/4$, where $E_{K\alpha}$ is the
\fe\ \ka\ line energy, there will be net downscattering. Therefore,
these wings could be interpreted as some form of 'hot Compton shoulder'.

\cite{owocki01} have computed profiles for X-ray lines emitted in the
winds of OB stars. They show clearly a characteristic red wing
whenever they are produced within the Wind-shock paradigm (see Fig. 4
of that reference), while they tend to show symmetric profiles if the
origin is coronal and, thus, produced at or very near the surface of
the star. Therefore, the characteristic profile of the lines in
Fig. \ref{fig:asym} suggests that the \fe\ \ka\ line is formed
in the shocked wind of hot stars illuminated by the X-ray source.

\section{Discussion}

\begin{figure}
\includegraphics[angle=-90,width=\columnwidth]{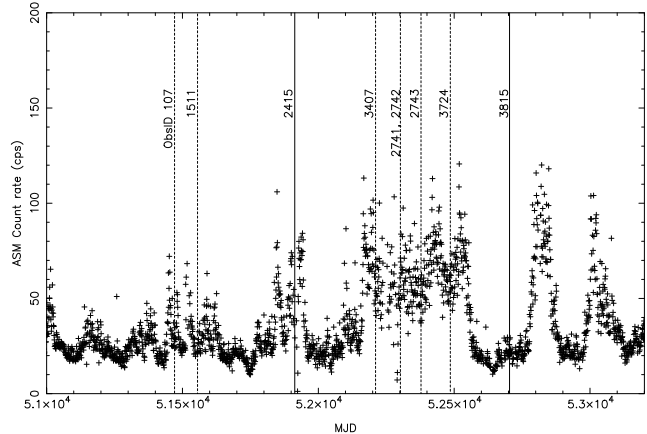}
\caption{{\it RXTE-ASM} lightcurve of Cyg X-1. Vertical lines represent the dates of the \chandra\ pointings included in this survey. Numbers along the lines refer to the corresponding ObsID. Solid lines correspond to positive detections of \fe\ \ka\ lines while dashed lines correspond to non detections.}
\label{fig:cygx1_asm}
\end{figure}


Our finding that narrow Fe K lines are present in every HMXB but are very
rare in LMXBs is in stark contrast with an early study by 
\cite{gott95} and a more recent study by \cite{asai00}.
While in the latter study over about 50$\%$ of line detections were
clearly pointing towards the existence of ionized Fe K lines,
many detections allowed to be interpreted as Fe line fluorescence.
Our analysis, on the other hand, shows that while $\sim 100\%$ of HMXBs
show a presence of a narrow line, only less than $\sim 10\%$ of LMXBs do.
Likewise, we do not find much Fe K edge absorption in LMXBs. This 
supports our previous findings and also the conclusion by 
\cite{miller09} that absorption in 
the interstellar medium dominates the neutral column density
observed in LMXB spectra.  

Studies of Fe fluorescence in X-ray binaries are not trivial and 
issues regarding statistics, bandpass, and competing physical
processes play a vital role in the interpretation of the results.
In general, the process of Fe line fluorescence is bandpass limited
to the Fe K line regions
unless secondary competing effects exist. This has the advantage that we do 
not need to deal with broad band binary modelings, which are generally
very difficult, uncertain, and even controversial. Our survey, however,
can very well discern moderately broadened line emission from neutral 
and warm Fe K fluorescence (1.94 \AA\ to 1.90 \AA) and emission from 
highly ionized species of Fe XXV (1.85 \AA) and Fe XXVI (1.78 \AA)
much to the level of the {\it ASCA} survey by Asai et al. (2000).

\cite{shapos09} found a significantly broadened Fe K line
with \emph{Suzaku}, at medium spectral, resolution with a line width exceeding
10$^4$ km s$^{-1}$, in the spectrum of Cyg X-2,
which was modeled by either a relativistic disk reflection model \citep{fabian82, laor91} or
an extended red-skewing wind model \citep{laurent07}. The \emph{Chandra} observation
\citep{schulz09} also shows such a broad line, but in contrast
resolves the line into the Fe XXV He-like triplet and the Fe XXVI H-like
line component ruling out Fe line fluorescence. Clearly, extreme 
cases of broad line emission as detected in some LMXBs (Bhattacharyya
\& Strohmayer 2007, Cackett et al. 2008) are difficult, if not impossible,
to detect with local continua. However, besides the fact that they are 
still controversial,
they also seem not to be associated with the Fe K edge absorption as the line fluorescence in our survey.

Our study shows that the narrow \fe\ K lines detected are produced in
the relatively cold wind of massive stars by reprocessing the X-ray
photons from the X-ray source. The reprocessing material is
spherically distributed around the X-ray source leading to narrow
\fe\ka\ line widths of $FWHM\lesssim 0.005$ \AA. For the measured
\fe\ka\ wavelengths this is equivalent to a velocity $v=c\frac{\Delta
  \lambda}{\lambda}\approx \frac{FWHM}{\lambda}\approx 770$
km/s. Assuming a typical \emph{beta}-law, for the velocity of the wind

\begin{displaymath}
v_{w}=v_{\infty}(1-R_{*}/r)^{\beta}
\end{displaymath}

with $\beta = 0.8$ and $r$ of the order of $2R_{*}$, we have
$v_{w}\sim 0.6v_{\infty}$. Typical terminal velocities for OB stars
are of the order of $v_{\infty}=1500$ km/s which yields $v_{w}\simeq
850$ km/s, slightly larger, but of the order of, the width of the
observed lines. Thus, the origin of the narrow component is compatible with the stellar wind. Systems with strong winds will present a strong (narrow)
\fe\ K line. It is not present in LMXB except in those rare cases
where the system presents a substantial wind component. Indeed,
4U1822$-$371 is the prototypical Accretion Disk Corona source, Cir X-1
presents a hot accretion disk wind \citep {schulz02_b}, GX1$+$4 is a symbiotic binary whose MIII donnor has a powerful wind and Her X-1 has an A type donor, the earliest of the late type companions. 

The wind however, can not be the only origin for narrow lines. Indeed, illuminated disks can produce narrow lines, as well, under the right conditions. It has been observed, for example, in the Seyfert 1 galaxy NGC 3783 originating, at least partially, from a relativistic accretion disk \citep{yaqoob05}. Since we do not observe this line in LMXBs, either their disks are too hot or not illuminated.

A particularly striking case is that of the HMXB Cyg X-1 where only two, out of nine observations analyzed, have shown \fe\ \ka\ in emission. In these two cases the emission is amongst the weakest of the whole sample. In Fig. \ref{fig:cygx1_asm} we show all the \chandra\ pointings overplotted on the {\it RXTE-ASM} lightcurve. As can be seen, the two observations where the line has been observed, marked by solid lines, do not correspond to the same spectral state of the source. During the ObsID 2415 observation, the source was in an intermediate state, with $\sim 50$ cps in the {\it ASM}. On the other hand, during the ObsID 3815 observation, the source was clearly in a low (hard) state, with $\sim 25$ cps. The lack of detection in the highest states (2741, 2742, 2743, 3407, 3724) could be explained by the Baldwin effect discussed in this paper: the increase of the luminosity of the X-ray source, increases the degree of ionization of the neutral Fe, thereby decreasing the strength of the fluorescence line. This would be supported by the findings by \cite{hanke09} and \cite{juett04} in which the neutral column densities decrease with increasing luminosity. The lack of detection during ObsIDs 107 (intermediate-low state) and 1511 (low state), can not be explained in this way and is enigmatic. As stated before, \citep{dunn08} have shown that, for the case of GX339$-$4, the Baldwin effect seems to be only effective in the high soft state and not in the low hard state. 

Chandra HETG spectra thus proves to be uniquely qualified to separate
narrow Fe K line fluorescence from hot and broad line components. Even
though HETG spectra detect far less hot Fe K lines in LMXBs than the
survey conducted by Asai et al. (2000), this can easily be explained
by the intrinsic variability of the the hot component in these sources
as well as fitting biases in CCD spectra.


\section{Conclusions}

We have reprocessed and analyzed the HETG spectra of all X-ray
binaries publicly available at the \chandra\ archive with specific
focus on the \fe\ K line region. The following conclusions can be
drawn from this analysis:

Fe $K\alpha$ fluorescence emission seem to be ubiquitous in
HMXBs. This emission varies throughout time for a specific source. In
particular, Cyg X-1 shows a rather weak emission, only in two out of
nine \chandra\ observations analyzed here. This emission is detected in two different spectral states: during a low hard state (ObsID 3815) and an intermediate state (ObsID 2415).  It vanishes for brighter (softer) states. A possible explanation could reside in the X-ray Baldwin effect studied before. Only in low luminosity states there remains a significant fraction of near neutral Fe, although much less than in other binaries of this survey. This remaining neutral Fe, however, becomes ionized in brighter states with the disappearance of the fluorescence line. However, the lack of detection during other intermediate (ObsID 107) and low hard states (ObsID 1511) defies this explanation and means that this can not be the only mechanism at work in this system.

In contrast, such emissions are found to
be very rare amongst LMXBs. Only four sources, out of 31 analyzed in this work, display the narrow component of Fe
$K\alpha$ in emission: 4U1822-37, GX1+4, Her X-1, and Cir X-1. This lack of narrow iron line is always accompanied by the lack of any detectable K edge. 
This finding, strongly suggests, that the neutral absorption column in LMXB is dominated by the interstellar medium while in HMXB the local absorption is very significant.

This
finding is in contrast with the previous work by \cite{gott95}, based
on spectra of lower resolution, where the majority of sources showing
Fe line emission were LMXB. Our findings, however, do not contradict
the ASCA study of Asai et al 2000 which claim the detection of hot
lines in the majority of LMXB. We point out though, that these hot
lines are not as frequently observed in \chandra\ HETG spectra.

The curve of growth ($EW(K \alpha)$ vs $N_{\rm H}$) is fully
consistent with a spherical distribution of reprocessing matter, formed
by cold near neutral Fe, around the X-ray
source. This reprocessing material must be close to the X-ray
source, as the line and continuum variations are closely correlated.
Those few LMXB displaying Fe $K\alpha$ in emission follow the same
correlations found for the HMXB.  Since the nature of the winds is
very different in LMXB and HMXB this means that the circumsource
material has lost all 'memory' of the donor. This is consistent with a
reprocessing site location very close to the compact object.

 We observe a moderate anticorrelation between $EW$ and the $L_{\rm
  X}$ of the source, on average. Some sources follow individually this
trend, over timescales from days to months, while other do not. This 'X-ray Baldwin effect' is reported here
for the first time, for the XRBs as a class. The immediate
interpretation is the increase in the degree of ionization of Fe with
the increasing $L_{\rm X}$ of the illuminating source which produces a concurrent
decrease in the efficiency of the fluorescence process.

We observe a Compton shoulder in the supergiant HMXB X1908$+$075 formed by
single Compton scattering of primary \ka\ photons. Together with the hypergiant system GX 301-2,
where such a shoulder has first been reported~\citep{watanabe03},
they form the very small group of galactic sources with such a feature detected. Other sources (LMC X-4, 4U1700$-$37) show asymmetric wings,
with the blue wings rising sharply and red wings declining
progressively. This effect can be explained as 'hot Compton
shoulders'.

Fe line fluorescence is produced in the stellar wind of massive stars. Systems with a substantial wind
component will show this line. Therefore it can be naturally observed
in all HMXB. On the other hand, donor stars in LMXB are not significant sources of stellar winds and other sites must be invoked for the origin of this line. Relativistic accretion disks, as observed in some Seyfert galaxies, might be suitable candidates for those few LMXB with positive detections. However, the lack of the narrow Fe line in the spectra of the vast majority of LMXB might be an indication that such accretion disks are too hot or not illuminated. 







\acknowledgments

We thank Julia Lee for making the LMC X-4 data available to us
previous to publication. JMT acknowledges the support of the Spanish
Ministerio de Educaci\'on y Ciencia (MEC) through the grant PR2007-0176, and the  MICINN through grants AYA2008-06166-C03-03 and Consolider-GTC CSD-2006-00070.

















\end{document}